\documentclass[journal]{IEEEtran}
\ifCLASSINFOpdf
\else
   \usepackage[dvips]{graphicx}
\fi
\usepackage{url}
\usepackage{graphicx}
\usepackage{amsmath,bm,bbm}
\usepackage{amssymb}  
\usepackage{dsfont}   
\usepackage{cite}        
 \usepackage{epstopdf} 
  \usepackage{xcolor} 
  \usepackage{subfigure} 
\usepackage{amsthm}    
\usepackage{enumerate}
\usepackage{makecell}  
\usepackage{diagbox,multirow }  
\usepackage{color,colortbl} 
\usepackage[ruled]{algorithm2e}  
\usepackage{algorithmic}

\usepackage{booktabs}
\usepackage{soul}

\usepackage{setspace}

\begin{document}

\title{Environment  Semantics Aided Wireless Communications: A Case Study of  mmWave Beam Prediction and Blockage Prediction}
\author{Yuwen Yang,  Feifei Gao, Xiaoming Tao, Guangyi Liu and Chengkang Pan
\thanks{Y. Yang and F. Gao are with  Institute for Artificial Intelligence Tsinghua University
(THUAI), State Key Lab of Intelligent Technologies and Systems, Beijing National Research Center for Information Science and
Technology (BNRist), Department of Automation, Tsinghua University, Beijing,
100084, P. R. China (email: yyw18@mails.tsinghua.edu.cn, feifeigao@ieee.org).}
\thanks{X. Tao is with the Department of Electronic Engineering, Tsinghua University, Beijing, P.R. China, 100084 (email:
taoxm@tsinghua.edu.cn).}
\thanks{G. Liu and C. Pan are with the China mobile communication research Institute, Beijing, China (e-mail: liuguangyi@chinamobile.com; panchengkang@chinamobile.com).}
}

\markboth{XXXX, VOL. XX, NO. XX, XXX }
{Shell \MakeLowercase{\textit{et al.}}: xxxx}
\maketitle
\vspace{-9mm}
\begin{abstract}
In this paper, we  propose an  environment semantics aided  wireless communication framework   to   reduce the transmission latency and improve the transmission reliability, where semantic information is   extracted from environment image data,   selectively encoded based on its task-relevance, and  then fused to make decisions for  channel related  tasks.
As a case study, we   develop an  environment semantics aided  \emph{network architecture} for mmWave communication systems, which is composed of a semantic feature extraction network, a feature selection  algorithm, a task-oriented encoder, and a decision network. With images taken from street cameras  and user's identification information as the inputs, the environment semantics aided  network architecture is trained to  predict the optimal  beam index and the blockage state for the base station.
It is seen that  without  pilot training or the costly beam scans, the environment semantics aided network architecture   can realize extremely efficient  beam prediction and   timely blockage prediction, thus meeting requirements for ultra-reliable and low-latency communications (URLLCs).
Simulation results demonstrate that compared with   existing   works, the proposed environment semantics aided  network architecture can reduce system overheads such as  storage space  and computational cost while achieving satisfactory prediction accuracy and protecting  user privacy.

\end{abstract}
\vspace{-2mm}

\IEEEpeerreviewmaketitle

\section{Introduction}\label{secintorf}
As identified by  Shannon and Weaver \cite{6773024},  communications can be categorized into three levels:
i) transmission of symbols;
ii) transmission of semantic information behind transmitted symbols;
iii) effectiveness of semantic information interaction.
In 1950s, Shannon   focused on the first level and derived a rigorous mathematical theory of communications based on probabilistic models, yielding the famous  Shannon capacity limit \cite{thomas2006elements}.
In the past decades, wireless communication systems  based on the Shannon information theory  have evolved from the first generation (1G) to the fifth generation (5G). Various advanced technologies, including massive multiple-input multiple-output (MIMO) \cite{8354789}, mmWave  communications \cite{8443598}, and deep learning (DL) based communications \cite{8663966},    continuously  promote  the system capacity to Shannon capacity limit.
In particular, DL  has been recognized  as a promising candidate for beyond-5G (B5G) communications due to its   remarkable
capability of learning intricate inter-relationships hidden in massive data.
Great progresses have been achieved by DL in  physical layer communications such as  channel estimation \cite{9288911,8353153,8795533,9175003}, data detection \cite{8052521,9791486,9018199},
channel feedback \cite{8482358,guo2019convolutional,9136588}, and beamforming
\cite{9279228,liu2021deep,9112250}, etc.
By either modifying or replacing conventional communication modules,
these studies  exploit deep neural networks (DNNs) to improve  system accuracies or reduce  computational complexities.
Despite of continuously emerging DL based communication  applications,
the Shannon capacity limit is  still the insurmountable upper bound.

Driven by the ambitious goals of the sixth generation (6G) and empowered  by the development of artificial intelligence, a higher level communication paradigm, i.e., semantic  communications,   has been proposed very recently, which no longer focuses on accurately recovering the transmitted symbols  but concerns on  precisely recovering the meaning behind the transmitted symbols \cite{9679803,9530497}.  The term ``semantic information'' refers to the information/meaning associated with the source signals.
Semantic information exists in various signal modalities like   texts, speeches, images, videos, and even random variables  etc \cite{Kalfaee,MalikMalik,8100143,8844998}.
Studies on semantic communications in the area of computer vision (CV) and natural language processing (NLP) have attracted  great  attentions.
The paradigm of semantic  communications is to  extract semantic information,  transmit semantic information, and ensure semantic information  being correctly interpreted by the receiver.
Several recent works  \cite{9398576,9632815,9450827,9685667} have provided concrete examples of semantic text \cite{9398576,9632815},  speech \cite{9450827}, and image \cite{9685667} transmissions.
In \cite{9398576}, the authors propose a DL  based semantic communication architecture
for text transmission, where  the Transformer structure \cite{vaswani2017attention} is adopted in the semantic encoder/decoder.
In the training process,  the semantic accuracy and the system capacity are jointly
optimized,  while  in the testing process the  word-error-rate and the peak signal-to-noise ratio (PSNR) are used to measure the accuracy of the source information recovery.
The corresponding simulations demonstrate the superiority of the semantic architecture over  traditional approaches.
Following the semantic communication architecture  in \cite{9398576}, the work \cite{9632815}  introduces an adaptive circulation mechanism into the Transformer structure  \cite{vaswani2017attention}.
With the adaptive Transformer,
the  semantic communication system becomes more flexible to transmit sentences with different semantic
information and  exhibits better robustness over   channel conditions.
In \cite{9685667}, the authors propose a DL based image semantic coding algorithm under a rate-perception-distortion optimization framework, which can achieve the state-of-the-art visually pleasing reconstruction and semantic preserving performance in the extreme low bit rate case.

{In fact, all the existing semantics communication systems only utilize the semantics of source signals to enhance the information representation  power of the transmission bits, which can be  more precisely referred to as source-oriented semantic communications (SOSCs).} The SOSC systems aim to promote   the achieve rate reaching the Shannon capacity limit, which is consistent with the goal of enhanced mobile broad band (eMBB).
Meanwhile, how to realize  ultra-reliable and low-latency communications (URLLCs) is also an essential   topic for the implementation of an efficient practical system. Since mmWave bands have  abundant spectrum resources and  support wider  subcarrier spacing, mmWave communication systems can better meet low latency requirements in URLLCs.
However, mmWave communication system is inherently unreliable due to its sensitivity to blockage and high penetration loss. To explore mmWave systems in URLLCs, the acquisitions of channel, interference, and blockage information are critical, especially for  decisions like the beam switching or the base station  (BS)  hand-off.


Intrinsically,  channels are determined by the  wireless propagation environment, while the propagation environment can be captured  by the images/visions from the cameras in cars or streets.
Hence, one may extract the channel semantic information  from environment images, named as  environment semantics or  channel semantics, to help  channel related downstream tasks.
By exploiting  environment semantics, channel related downstream tasks could be accomplished without conventional channel estimation, which  can   significantly save the time and the computation cost related to the pilot training or the channel feedback, and  even  predict  burst channel interference  to improve the communication reliability.
Actually, some  works  have already  proposed to use environment images to help  channel related downstream tasks, e.g.,   beam selection \cite{weihua20192d}, channel covariance matrix estimation \cite{9523557}, and blockage prediction \cite{9512383}. These vision based works demonstrate that  channel information could be effectively attained from environment images to build   communication connections without pilot overheads.
However, original image usages for channel related tasks require expensive storage space, high computational cost, and huge energy consumption, which is unacceptable  for the base stations (BSs) or users.
Moreover,  environment image usages may   involve complex social issues such as user privacy, public safety, and management policies, especially when  the third party cameras (e.g., the ubiquitous surveillance cameras) are employed to provide  the environment images.
Compared with directly utilizing  environment  images, operating with   environment semantics  naturally protects  privacy. This is because that only the class and the layout information of objects are preserved in environment semantics, and the surface characteristics of objects are eliminated.
Furthermore, environment semantics is associated with various  objects, including key scatterers of  propagation channels and objects that are  irrelevant to propagation channels. By retaining  channel relevant semantics and mitigating channel irrelevant semantics, the system overheads such as storage space and computational cost could be also reduced.
The   environment semantics aided  communication system focuses on exploring channel semantics hidden in environment images, and therefore can be regarded as channel-oriented semantic communications (COSCs).

In this paper, we propose a framework for COSCs,
  where   environment semantics  is   extracted from environment images,  selectively encoded based on its  channel-relevance, and  then  fused to make decisions for specific  channel related downstream tasks.
To demonstrate the effectiveness of the  proposed framework, we develop an environment semantics aided network architecture for mmWave beam prediction and blockage prediction as a case study. The environment semantics aided network architecture is composed of an  environment semantics extraction network, a feature selection (FS)  algorithm,  a task-oriented encoder network, and a decision network.
With  images taken from street cameras, the  environment semantics is obtained, selected,  encoded, and then transmitted to BS.
After fusing the encoded  channel    semantics and identification information of the target user, BS can obtain the predicted beam index and blockage state by the decision network.
The environment semantics aided beam prediction can realize extremely low-latency without  pilot training or the costly beam scans, while the environment semantics aided blockage prediction can support  ultra-reliable communication links. Therefore, the proposed  environment semantics aided   framework offers  extraordinary  application values for URLLCs.
Simulation results based on the autonomous driving  and  the ray-tracing channel simulators demonstrate the effectiveness of the proposed environment semantics aided  communication framework, which further    validates its great application potential in URLLCs.



 The rest of this paper is organized as follows.
The frameworks of COSCs and SOSCs are proposed in Section~\ref{secsena}.
As a case study, the environment semantics aided network architecture for mmWave beam prediction and blockage prediction is developed in Section~\ref{seccase}. Numerical simulations  are provided in Section \ref{secsimu}. The main conclusions are
given in Section \ref{secconcul}.

\emph{Notation:}
The bold and lowercase letters denote vectors;
 the notation  $(\cdot)^T$ denotes the transpose   of a matrix or a vector;
the notations ${\mathds{C}^{m\times n}}$ and ${\mathds{R}^{m\times n}}$  represent  the $m\times n$ complex and the $m\times n$  real vector space, respectively;
the notation $\left\| {\bm x} \right\|_2$  denotes the $L_2$  norm of  $\bm x$;
the notation $| {\bm x} |$  denotes the $L_1$  norm of  $\bm x$ while    $| X|$  denotes the size  of  the set $X$.



\section{Source Semantics and Channel Semantics}\label{secsena}

In this section, we will introduce
 SOSCs,
and then propose a framework of  environment semantics aided  communication systems as a representative instance of
 COSCs.
A communications system working with both SOSCs and COSCs will be named as a generalized semantics communications systems. We will then heuristically analyze the source semantics and channel semantics.
\subsection{Existing Semantic   Communication Systems: SOSCs}
\begin{figure*}[!t]
\centering
\includegraphics[width=160mm]{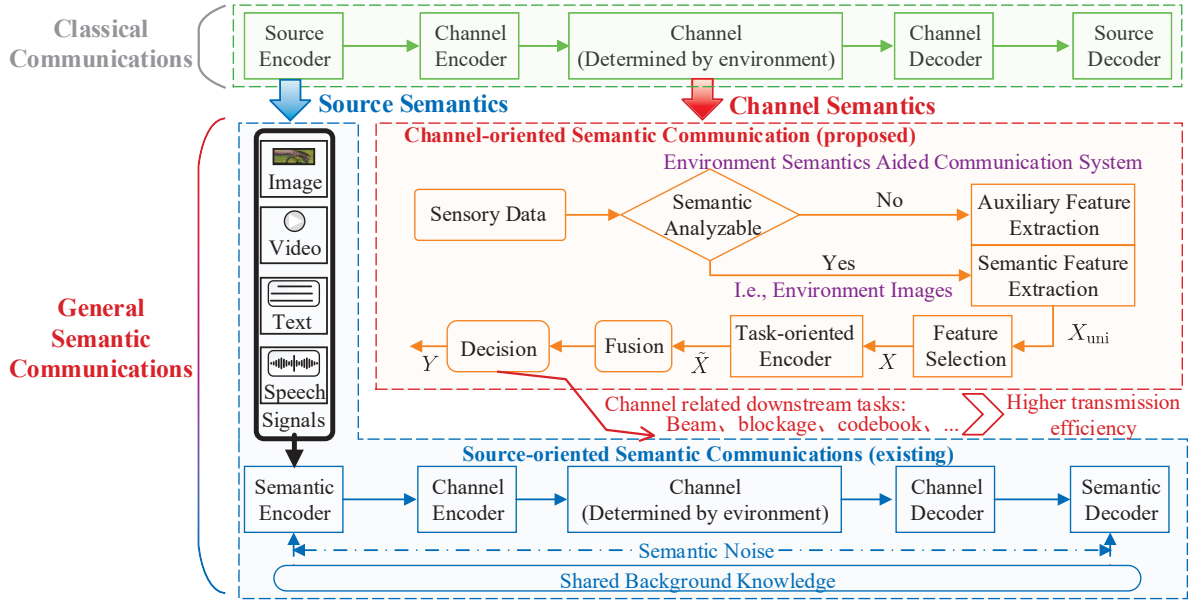}
\caption{Frameworks of a  classical (the upper part), an existing semantic (the bottom part), and an environment semantics aided (the middle part) communication systems. The  environment semantics aided    and the existing semantic communication systems are the instances of  COSCs and  SOSCs, respectively.   }
\label{figsemantic}
\end{figure*}

Fig.~\ref{figsemantic}  illustrates the frameworks of a classical (the upper part) and an existing semantic (the bottom part)  communication systems.
In a classical communication system, signals with various modalities are first compressed by a source encoder to save transmission cost.  Then, the designed redundancies are added by a channel encoder  to combat  physical channel
noises in transmission links, and the received signals experience the inverse process in the receiver side.
Unlike  classical communications whose goal is to ensure each transmitted bit being correctly received, the
 SOSC systems \cite{9398576,9632815,9450827,9685667} aim to guarantee the meaning behind the transmitted bits, i.e., the semantic information of the transmitted signals,  being  correctly understood by the receiver.
Details about the modules of the existing semantic communication systems are given in the following.

\noindent \textbf{Semantic encoder}    extracts    semantic features from the source signals and  compresses them to save the transmission cost. Next, the compressed semantic features go through a channel
encoder  and are transmitted over physical channels.

\noindent \textbf{Semantic decoder} interprets  the received semantic features and recover  the original source signals, i.e., conducting  the inverse process of the semantic encoder.

\noindent \textbf{Background knowledge base (BKB)} is the prerequisite of semantic communication systems, which can be independently established or interactively  shared by the transmitter and the receiver. The semantic encoder  extracts semantic features based on the background knowledge at the transmitter, while the semantic decoder interprets semantic features and restores the  original source signals based on the background knowledge at the receiver.
The BKBs can learn from previous signal data and periodically update with the gradual progress of communications.
Naturally, BKBs are different in various contexts. For example, high-frequency words in  business and medical scenarios are different, and therefore the corresponding BKBs contain  different semantic corpora. Moreover,  BKBs should have different formats for signals in various modalities, e.g., text, image, and speech, etc.
Besides, BKBs at the transmitter and the receiver might be different due to cognitive bias or belated sharing.

\noindent \textbf{Semantic noise} comes from  semantic ambiguity of received signals or mismatches between the BKBs at the transmitter and the receiver, resulting in semantic errors in the interpretation and recovery processes. Note that the physical channel noise in transmission environments might lead to symbol errors of the received signals, which can also cause semantic errors.

\subsection{Environment  Semantics Aided Communication System: COSCs}

 The framework of the environment semantics aided communication system is  displayed in the middle part of Fig.~\ref{figsemantic}, where    semantic information is extracted from environment images for channel related downstream tasks, e.g.,   beam prediction,  blockage prediction, and codebook designs, etc.
We refer the semantic information extracted from environment images as environment semantics or the channel semantics, which naturally protects user privacy by eliminating the surface characteristics of objects and preserving only the class  and the layout information of objects.
Since the environment semantics aided communication system exploits the channel semantics hidden in the environment images to assist channel related downstream tasks, it can be regarded as an instance of COSCs.

To illustrate  the   idea, suppose the goal of environment semantics aided communication systems is to predict the parameter $Y$ from a set of multimodal observations/samples, including environment images \cite{9277535}. Here $Y$ can be  the beam index/vector  in beam prediction tasks, the blockage flag in  blockage prediction tasks, the codebook in codebook design tasks, or the channel covariance matrix in channel covariance matrix estimation tasks, etc.
We refer   features extracted  from other relevant modalities as \emph{auxiliary features}, which can   help to identify target users or provide extra system information.
Denote the  universal set of the extracted features as $X_{\textrm{uni}}$.
By utilizing FS methods, the most efficient  feature set
could be selected from $X_{\textrm{uni}}$.
Then, the task-oriented encoder maps $X$ into $\tilde{X}$ to save system overheads  (e.g., storage space, transmission delay, and energy consumption, etc).
The compressed semantic feature set $\tilde{X}$ is sent to the fusion module  for further decisions in channel related downstream tasks.

\subsection{Analyses of Source Semantics and Channel Semantics}
Signals and channels are the two   major components of communication systems.
Existing  semantic communication systems  \cite{9398576,9632815,9450827,9685667} are SOSCs that focus on the semantics of the source signals to enhance the representation capacity of transmitted bits. The proposed environment semantics aided communication system belongs to COSCs that extract the  channels semantics  from multimodal data like environment   images to assist the accomplishment of channel related downstream tasks without  pilot training.
In other words, SOSCs improve the representation capacity of transmitted bits by utilizing the  source signal semantics, i.e., pursuing the goal of eMBB.
In contrast, COSCs improve the transmission efficiency by  exploiting the channel semantics hidden in multimodal data, i.e., targeted at the requirements of URLLCs. In the environment semantics aided communication system,  channel semantics refers to the shape and classes of channel relevant scatterers, which can be obtained   from the environment images.
Besides, the proposed  environment semantics aided communication system
is a concrete realization of task-oriented  communications, where the module interactions are   carried out to accomplish a specific task while the  information not strictly relevant to the task would be  mitigated.  \cite{strinati20216g}.

\section{Case Study of  Channel Semantics: MmWave Beam Prediction and Blockage Prediction }\label{seccase}

In this section,  we   develop an environment semantics aided network architecture for mmWave beam prediction and blockage prediction to meet requirements of URLLCs as
a case study.

\subsection{System Model}
Consider a massive MIMO system, where a BS is equipped with  $N_t\!\gg\! 1$ antennas in the form of uniform linear array (ULA) and serves multiple  single-antenna users.
The received frequency domain signal of a  user on the $k$-th subcarrier  is
\begin{equation}\label{equreceive}
r[k]=\bm h ^{T}[k] \bm w  s[k] + \varepsilon [k],
\end{equation}
where $r[k]\in \mathds{C}^{1\times 1}$ is the received signal, $ s[k]\in \mathds{C}^{1\times 1}$ is the transmit signal, $\bm w \in \mathds{C}^{N_t\times 1} $   is the transmit beamforming vector,    and
$  \varepsilon[k]\!\!\sim\!\! \mathcal{CN}\left(0, \sigma_{\varepsilon}^{2} \right) $  is the additive white Gaussian noise. 
Moreover,  $\bm h [k]\in \mathds{C}^{M\times 1}$ is the downlink channel that can be written as \cite{alkhateeb2019deepmimo} 
 \begin{equation}\label{equchannel}
\bm h [k]  = \sum\nolimits_{l}  {\alpha _{ l}{e^{ - j2\pi f_{\textrm{D},k}\tau_{ l} + j{\phi _{ l}}}} \bm a\left( {\theta_{\mathrm{az}}^{ l}, \theta_{\mathrm{el}}^{ l}} \right)},
\end{equation}
where  $f_{\textrm{D},k}$ is the  frequency of the $k$-th downlink subcarrier, while  $\alpha _{ l}$, $\phi _{ l}$, and $\tau _{ l}$ are the  attenuation, phase shift, and  delay
 of the $l$-th path, respectively. In addition, $\bm a\left( {\theta_{\mathrm{az}}^{ l}, \theta_{\mathrm{el}}^{ l}} \right)$ is the array manifold vector\footnote{When other types of  antenna arrays are adopted, the array manifold vector should be changed accordingly. Note that the proposed network architecture is not limited by   specific  antenna array shape, and therefore is applicable for  array with arbitrary geometry.} defined as
\begin{eqnarray}\label{equavec}
\bm a\left( {\theta_{\mathrm{az}}^{ l}, \theta_{\mathrm{el}}^{ l}} \right){\kern -8pt}&={\kern -8pt}&\left[1, e^{j\varpi \sin \left(\theta_{\mathrm{el}}^{ l}\right) \cos \left(\theta_{\mathrm{az}}^{ l}\right)}, \ldots\right.  \nonumber \\
& &{\kern 5pt}\left.\ldots, e^{j \varpi\left(M-1\right) \sin \left(\theta_{\mathrm{el}}^{ l}\right) \cos \left(\theta_{\mathrm{az}}^{ l}\right)}\right]^{T},
\end{eqnarray}
 where  $\varpi ={{2\pi d f_{\textrm{D},k}}}/{c }$, $d$  is  the antenna spacing,  $c$ is the speed of light, and $\{\theta_{\mathrm{az}}^{ q}, \theta_{\mathrm{el}}^{ l}\}$  is the \{azimuth, elevation\} angle  of  arrival.
 Assume the beamforming vector $\bm w  $  is selected from  the beam codebook  $\mathcal{W}$, where $M_{\textrm{bm}}=|\mathcal{W}|$ is the codebook size.
 The achievable transmission rate corresponding to the beam vector ${\bm w}$ can be written as
 \begin{eqnarray}\label{equarate}
\textrm{Rate}_{\bm w}=  \frac{1}{K} \sum_{k=1}^{K} \log _{2}\left(1+\frac{P_{k}}{\sigma^{2}}\left|\bm h ^{T}[k] \bm w\right|^{2}\right),
\end{eqnarray}
where $P_{k}=E[\left|s [k]\right|^{2}]$ is the power of the transmitted signal\footnote{To simplify the notation, we drop the sub-carrier index $k$
in the rest of the paper, e.g., replacing  $\bm  r[k], \bm h_{u}(f_{\textrm{D},k})$, and $\bm s[k]$
with $\bm  r, \bm h(f_{\textrm{D}})$, and $\bm s$, respectively.}.
 The optimal beamforming vector can be obtained by maximizing the transmission rate, i.e.,
 \begin{eqnarray}\label{equaraterhh}
\bm{w}^{\textrm{opt}}=\underset{\bm w \in \mathcal{W}}{\arg \max }\  \textrm{Rate}_{\bm w}.
\end{eqnarray}
In fact,  the beamforming vector $\bm w  $ is designed to  concentrate the transmitter power  on a narrow beam to compensate for the  high penetration loss of mmWaves. To obtain the optimal beamforming vector $\bm{w}^{\textrm{opt}}$, conventional algorithms need to estimate  accurate channel information by extensive downlink pilot training, thus leading to unacceptable time delay especially for massive MIMO systems. Besides, exhaustively searching on the a large codebook also greatly increases the time delay, which makes it  impractical  to apply   conventional algorithms   to URLLCs.
In the meantime, based on field measurements \cite{9689054}, the channel power is generally  concentrated in the  first few paths, especially the line-of-sight (LOS) path.
Therefore,   LOS link blockages caused by obstacles would significantly reduce  the achievable transmission rate, and even lead  to  link disconnections. To ensure the link reliability of mmWave communications, the prediction of future link blockages is critical to   beam/BS switching decisions.

 \begin{figure*}[!t]
\centering
\includegraphics[width=160mm]{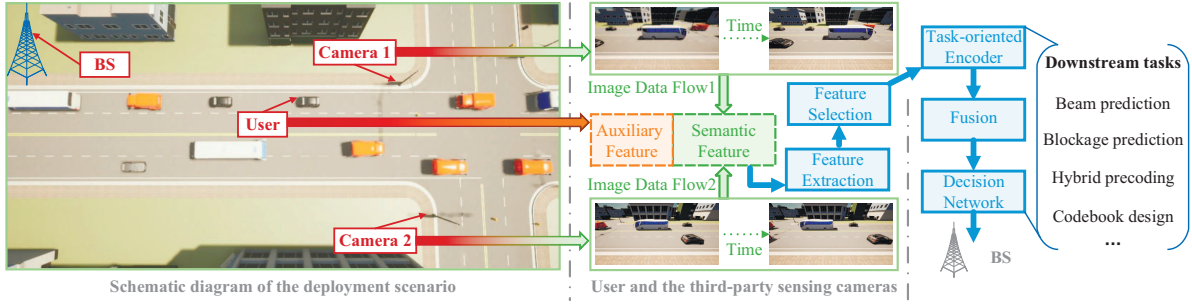}
\caption{The overall diagram of the environment semantics aided communication system. Here two cameras are schematically installed in the deployment scenario. In fact, the number and positions of the cameras can be set according to practical conditions and demands, which will not affect the algorithm design. }
\label{fignetmodel}
\end{figure*}

\subsection{Problem Formulation}


As indicated in Eq.~\eqref{equchannel}, millimeter  massive MIMO channels have sparse  structures associated with the parameters such as attenuations, phase shifts, and delays, etc. Naturally, these channel
parameters can be regarded as one type of  semantic information.  However, from an  intrinsical perspective, wireless channels are determined by the propagation paths or more  straightforward by the key scatters in propagation environment, e.g., vehicles or roads. Therefore, there exists channel related semantic information in corresponding environment images that  could be extracted to improve the channel related  tasks, e.g., the beam prediction,  blockage prediction, hybrid precoding, and codebook design, etc.

Suppose that cameras, i.e., the sensing devices  installed on both sides of  streets, can capture image data flows to assist BS in   channel related  tasks. As shown in Fig.~\ref{fignetmodel},
different cameras can provide image data from different perspectives, thus avoiding  the target user  being blocked by other vehicles nearby.
Since images from cameras are municipal infrastructure data,  which involve complex social  issues such as user privacy, public safety, and management policies, etc, they should not be directly sent to the fusion center for communication services.
By extracting environment semantics from images, the  surface characteristics of buildings, vehicles, or roads, etc, are eliminated, while  only  the class  and the layout information of objects are preserved. Therefore, environment semantics is naturally more private than original images, and facilitates potential new privacy transmission protocols.
Meanwhile, user's identification information,  such as locations, historical beam indices, or  other predetermined information that could be used to distinguish users, should also be sent to the fusion center in a secure and private way\footnote{If restricted, the feature extraction could be independently implemented by  terminal devices without communications with subsequent communication tasks. More details about how to strictly protect user privacy involve  specific device transmission protocols and encryption methods, which are possibly left as future work. }. 
 We refer the features extracted from  users' identification information and environment semantics from cameras as auxiliary and semantic features, respectively.
Raw  multimodal data from communication or sensing devices are first transformed into auxiliary or semantic features, and then selectively encoded by the task-oriented encoder.
After fusing the  encoded features from  the target user and  the  cameras on streets,   optimal decisions can be obtained by the designed network at  BS.

\begin{figure*}[!t]
\centering
\includegraphics[width=165mm]{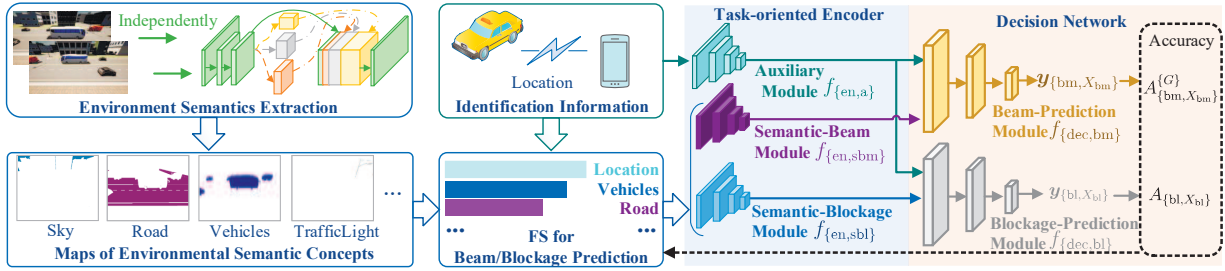}
\caption{Diagram of the environment semantics aided network architecture for beam prediction and blockage prediction.}
\label{fignetwork}
\end{figure*}

\subsection{Environment Semantics Aided Network Architecture for Beam Prediction and Blockage Prediction}

Fig.~\ref{fignetwork} displays the diagram of the environment semantics aided network architecture for beam prediction and blockage prediction, consisting of the environment semantics extraction network, the FS algorithm, the task-oriented encoder, and the decision network.
Specifically, the environment  semantics  extraction network  is deployed independently at each camera to extract the environment  semantics from the corresponding environment images.
The output of the environment  semantics  extraction network is the segmentation map, i.e., the pixel-wise category labels of the environmental semantic concepts (e.g.,   ``road'', ``vehicle'', ``trafficlight'',  and ``sky'', etc).
After multiplying the segmentation map by the corresponding  zero-masks, one can obtain  maps of various environmental semantic concepts, as shown in the lower left quarter of Fig.~\ref{fignetwork}.
We refer a map of each environmental semantic concept  as an instance of  one semantic feature.
The  universal feature set $X_{\textrm{uni}}\buildrel \Delta \over =\{X_{\textrm{uni},v}| i=1,\cdots,V_{\textrm{uni}} \} $ contains both the semantic features from environment images and the auxiliary features from  users' identification information, where $V_{\textrm{uni}}=|X_{\textrm{uni}}|$ denotes the size of $X_{\textrm{uni}}$.
Note that many types of user's identification information, such as  locations,  historical beam indexes, or other predetermined information, could all be added to $X_{\textrm{uni}}$ for further selection.  In this work, we adopt the location as an example.
Then, we utilize the FS algorithm to select the optimal feature sets, i.e., \[X_{\textrm{bm}}\buildrel \Delta \over =\{X_{\textrm{bm},v}| i=1,\cdots,V_{\textrm{bm}},X_{\textrm{bm},v}\in X_{\textrm{uni}}  \} \] and \[X_{\textrm{bl}}\buildrel \Delta \over =\{X_{\textrm{bl},v}| i=1,\cdots,V_{\textrm{bl}},X_{\textrm{bl},v} \in X_{\textrm{uni}}  \},\] from $X_{\textrm{uni}}$ for  beam prediction and blockage prediction, where   ${V_{\textrm{bm}}}$ and ${V_{\textrm{bl}}}$ are the sizes  of  $X_{\textrm{bm}}$ and $X_{\textrm{bl}}$, respectively.
Since the user's  identification information  is necessary to distinguish the target user from other possible users,
the user location  would definitely be retained by the FS algorithm as the auxiliary feature for both beam prediction and blockage prediction\footnote{When more types of users'  identification information are involved, which  type  of users'  identification information is retained  depends  on the FS algorithm.}.   For clarify, we denote the location as the feature   $X_{\textrm{bm},1} \buildrel \Delta \over = X_{\textrm{bl},1}$. Hence, $\{X_{\textrm{bm},v}\}_{v=2}^{V_{\textrm{bm}}}$ and $\{X_{\textrm{bl},v}\}_{v=2}^{V_{\textrm{bl}}}$ are the semantic features selected from  the environment  semantics for beam prediction and blockage prediction,  respectively.
The task-oriented encoder is composed of  the auxiliary, the semantic-beam, and the semantic-blockage modules.
The decision network includes the blockage-prediction and the beam-prediction modules.
With $X_{\textrm{bm}}$  as the input, the three  modules, i.e., the auxiliary,  the semantic-beam, and the beam-prediction modules, are jointly trained to predict the beam.
With $X_{\textrm{bl}}$  as the input, the three  modules, i.e., the auxiliary,  the semantic-blockage, and the blockage-prediction modules, are jointly trained to predict the blockage.
Since pilot training and costly beam scans are unnecessary, the environment semantics aided network architecture  is expected to realize   low-latency beam prediction and blockage prediction to ensure efficient and reliable mmWave communication links, so as to fulling the goal of URLLCs.


In the following, we will  present  the environment semantics extraction network, followed by the structures of above-mentioned modules.
Then, the FS algorithm for  beam prediction and blockage prediction are presented respectively.
Lastly, the overall training steps of the environment semantics aided network architecture  for beam prediction and blockage prediction are provided.

  \begin{figure*}[!t]
\centering
\includegraphics[width=150mm]{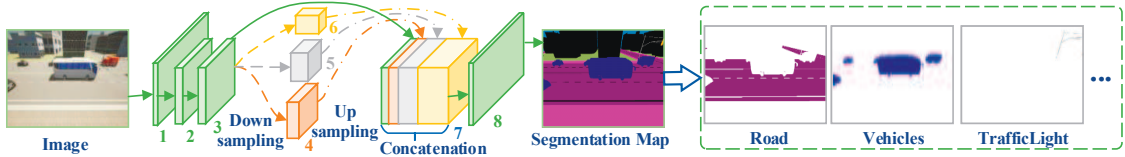}
\caption{Illustration of the environment semantics   extraction.  A green cuboid represents a feature map  output by a basic residue block. The number  next to the cuboid is the marked index of the feature map   but does not necessarily correspond to the actual number of blocks. Cuboids in other colors represent feature maps of different  size in the pyramid pooling module.
Notice that cuboids with smaller side areas typically are thicker in the horizontal direction. This is because that when images are down sampled, we tend to increase the filter number to avoid information loss, which is also a common trick  in computer vision.  }
\label{fignetseg}
\end{figure*}

\subsubsection{The environment  semantics  extraction network}
In semantic  coded transmission work \cite{9685667}, a semantic-perceptual loss defined by the semantic segmentation network PSPNet \cite{8100143} is adopted for image compression and reconstruction,  which is experimentally demonstrated to yield better reconstruction performance than common pixel-level losses at the same transmission bit rate.
The work \cite{9685667} also validates that the reconstructed images suffering less semantic information losses could achieve better performance in downstream tasks like object detection.
Inspired by this, we  utilize   PSPNet to  extract environment semantics from environment images, as shown in Fig.~\ref{fignetseg}. Inheriting from the well-known ResNet \cite{7780459}, PSPNet is composed of a serial of  basic residue blocks, and each residue block is sequentially stacked by several convolution,  batch normalization (BatchNorm), and ReLu   layers.
We refer the outputs of hidden layers as the feature maps, and mark them with serial numbers in Fig.~\ref{fignetseg} for  illustration convenience.
The main difference of PSPNet compared with ResNet is  the pyramid pooling module, where the $3$-rd feature map first
  goes through three pooling layers with different down-sampling rates in parallel.
Then,  the three down-sampled feature maps independently and  successively  go  through  a  convolution  and a BatchNorm layers,  obtaining three feature maps with different pyramid scales, i.e., the \{4,5,6\}-th  feature maps.
Next, the \{4,5,6\}-th  feature maps are up-sampled   by  bilinear interpolations such that they can be concatenated together with the   $3$-rd feature map as the final pyramid pooling global feature, i.e., the $7$-th feature map.
Note that the number of residue blocks, convolutional filters, and pyramid levels  can all be adjusted on demand. The principle of the pyramid pooling module is that  1) feature maps with smaller scales has larger sizes of receptive fields\footnote{The receptive field is the locations in deeper layers correspond to the locations in the input image they are path-connected to.};  2) layers with  larger receptive fields  could learn more about the global features of images while  layers with  smaller receptive fields could learn more about the local features of images;
3)   the pyramid pooling module could learn a comprehensive feature  of
 images by  fusing feature maps with varying scales.
Denote the   input image as $\bm g \in \mathds{R}^{3\times H\times W}$, where ``3'' denotes the RGB channels, $H$ is the image height, and $W$ is  the image width.
The label  and the  output of PSPNet are  $\hat{\bm c} \in \mathds{R}^{ M_{\textrm{con}}\times H\times W}$ and $\bm c \in \mathds{R}^{  H\times W}$, respectively, where  $M_{\textrm{con}}$ is the number of semantic feature categories, and the $(i,j)$-th entry of $\bm c$  is the  category label  of  the environmental semantic concept at the pixel coordinate $(i,j)$, i.e., $\bm c(i,j)=0,\cdots,M_{\textrm{con}}-1$.
The loss function of the PSPNet is given by
\begin{align}\label{equdentsemp}
f_{\{\textrm{Lpsp}\}}=- \frac{1}{HW}\sum_{i,j}\log \left(\frac{\exp (\hat{\bm c}(\bm c(i,j),i,j))}{\sum\nolimits_{m=0}^{M_{\textrm{con}}-1} \exp (\hat{\bm c}(m,i,j))}\right).
\end{align}

By using the ADAM algorithm to minimize $f_{\{\textrm{Lpsp}\}}$ until the convergence,
the segmentation map can be obtained by
\begin{equation}\label{equaccfptr}
\bm c_\textrm{map} = \underset{m=0,\cdots,M_{\textrm{con}}-1}{\arg\max} \ \hat{\bm c}(m,i,j).
\end{equation}\label{equaccpsp}
The   extraction accuracy of the environment semantics   is then given by
\begin{eqnarray}\label{equaccpsp}
A_{\textrm{PSP}}= \frac{\sum_{i,j}\sum\nolimits_{n=1}^{N_{\textrm{tot}}}\mathds{1}(\bm c_\textrm{map}^{(n)}(i,j) =\bm c^{(n)}(i,j))}{HWN_{\textrm{tot}}} ,
\end{eqnarray}
where the superscript ${(n)}$ denotes the $n$-th sample, $\mathds{1}$ is the indicator function, and $N_{\textrm{tot}}$ is the total number of testing samples.
The accuracy   $A_{\textrm{PSP}}$ measures the  accuracy of the environment semantics  extraction.
After obtaining the segmentation map,  i.e., $\bm c_\textrm{map}$, the  environment semantics such as   ``road'', ``vehicle'',
``trafficlight'', ``pedestrian'', ``sky'', and ``building'', etc, can be easily separated by multiplying zero-masks.

\subsubsection{The auxiliary module}

With the  location as  the input\footnote{In this example work, we can prejudge that the user location would be the input of the auxiliary-module.  When more types of users'  identification information are involved, the design of the auxiliary module would be more complex and depends on which type of users'  identification information  is chosen by the FS algorithm.}, the auxiliary module is composed of several fully connected (FC) blocks, and each FC block is sequentially  stacked by an FC,  a BatchNorm, and a ReLu  layers.
Denote the mathematical function of the auxiliary module as  $f_{\{\textrm{en,a}\}}$ for future use.
Although we adopt the same network structure of the auxiliary module for  both beam prediction and blockage prediction, the specific  network parameters of  the auxiliary module   would be trained independently for  beam prediction and blockage prediction.

\subsubsection{The semantic-beam module}
With the selected semantic features $\{X_{\textrm{bm},v}\}_{v=2}^{V_{\textrm{bm}}}$  as the input, the semantic-beam module is built by stacking several convolution-blocks, and each   convolution-block contains  a convolution,  a BatchNorm,   and a Relu  layers  in sequence.
The output size of the semantic-beam module can be changed by adjusting the number of convolution blocks, the number of filters in the convolution layer, or the down-sampling rate of the pooling layer.
Denote the mathematical function of the semantic-beam module as  $f_{\{\textrm{en,sbm}\}}$ for future use.

\subsubsection{The semantic-blockage module}
With the selected semantic features $\{X_{\textrm{bl},v}\}_{v=2}^{V_{\textrm{bl}}}$  as the input, the semantic-blockage module has similar structure with the semantic-beam module, i.e., stacked by several convolution blocks.
Denote the mathematical function of the semantic-blockage module as  $f_{\{\textrm{en,sbl}\}}$ for future use.
\subsubsection{The beam-prediction module}
The input of the beam-prediction module can be given as
 \begin{eqnarray}\label{equfusing}
\bm x_{\{\textrm{bm},X_{\textrm{bm}}\}}=[f_{\{\textrm{en,a}\}}(X_{\textrm{bm},1}), f_{\{\textrm{en,sbm}\}} (\{X_{\textrm{bm},v}\}_{v=2}^{V_{\textrm{bm}}})].
\end{eqnarray}
With  $\bm x_{\{\textrm{bm},X_{\textrm{bm}}\}}$ as the input, the beam-prediction module is composed of several FC, dropout, and Relu   layers. Define the  mathematical  function of the  beam-prediction module  as $f_{\{\textrm{dec,bm}\}}$.
Then, the output of the beam-prediction module is
\begin{eqnarray}\label{equdecf}
\bm y_{\{\textrm{bm},X_{\textrm{bm}}\}} =f_{\{\textrm{dec,bm}\}}(\bm x_{\{\textrm{bm},X_{\textrm{bm}}\}}).
\end{eqnarray}
The cross entropy loss is adopted for beam prediction, which is given by
\begin{align}\label{equdentrop}
f_{\{\textrm{Lbm},X_{\textrm{bm}}\}}=-\log \left(\frac{\exp (\bm y_{\{\textrm{bm},X_{\textrm{bm}}\}}[y_{\{\textrm{bm,lab}\}}])}{\sum_{m=1}^{M_{\textrm{bm}}} \exp (\bm y_{\{\textrm{bm},X_{\textrm{bm}}\}}[m])}\right),
\end{align}
where $y_{\{\textrm{bm,lab}\}}$ is the beam label.
The  auxiliary, the semantic-beam, and the  beam-prediction modules  are jointly trained by applying  adaptive moment estimation (ADAM)  algorithm \cite{kingmaadam} to minimize $f_{\{\textrm{Lbm},X_{\textrm{bm}}\}}$ until the convergence.
Denote the Top-$G$ beam-index set of the $n$-th testing sample  as $Y_{\{G,X_{\textrm{bm}}\}}^{(n)}$, which contains   the   indices  corresponding to $G$ maximum values of the output vector $\bm y_{\{\textrm{bm},X_{\textrm{bm}}\}}^{(n)} $.
Then, the Top-$G$ beam prediction accuracy can be obtained by
\begin{eqnarray}\label{equacckdcn}
A_{\{G,X_{\textrm{bm}}\}}=\sum\nolimits_{n}^{N_{\textrm{tot}}}\mathds{1}( y_{\{\textrm{bm,lab}\}}^{(n)}\in Y_{\{G,X_{\textrm{bm}}\}}^{(n)})/N_{\textrm{tot}}.
\end{eqnarray}
Denote the Top-$G$ beam-vector set as $\mathcal{W}_{\{G,X_{\textrm{bm}}\}}^{(n)} $, where each beam vector in  $\mathcal{W}_{\{G,X_{\textrm{bm}}\}}^{(n)} $ is selected from  the codebook   $\mathcal{W}$ according to the beam index set $Y_{\{{G},X_{\textrm{bm}}\}}^{(n)}$.
The  transmission rate  corresponding to ${Y_{\{{G},X_{\textrm{bm}}\}}^{(n)}}$  can be written as
 \begin{eqnarray}\label{equarate}
\textrm{Rate}_{Y_{\{{G},X_{\textrm{bm}}\}}^{(n)}} = \underset{\bm w \in \mathcal{W}_{\{G,X_{\textrm{bm}}\}}^{(n)} }{\max }\ \textrm{Rate}_{\bm w}.
\end{eqnarray}
Define the transmission rate ratio (TRR) of the Top-$G$ beam as the ratio between the  transmission rate corresponding to ${Y_{\{{G},X_{\textrm{bm}}\}}^{(n)}}$ and the optimal  transmission rate corresponding to the  beam $\bm{w}^{\textrm{opt}}$ in Eq.~\eqref{equarate}, i.e.,
 \begin{eqnarray}\label{equarate}
\textrm{TRR}_{\{G,X_{\textrm{bm}}\}} =\frac{1}{N_{\textrm{tot}}}  \sum\nolimits_{n}^{N_{\textrm{tot}}}  \frac{\textrm{Rate}_{Y_{{\{G,X_{\textrm{bm}}\}}}^{(n)}}}  {\textrm{Rate}_{\bm{w}^{\textrm{opt}}}}.
\end{eqnarray}
It should be mentioned that with  the Top-$G$ beam-vector  set   $\mathcal{W}_{\{G,X_{\textrm{bm}}\}}^{(n)} $, BS only needs to scan $G$ times to obtain the optimal beam. In this way, the environment semantics aided network architecture   offers an ideal initial beam search range, and thus significantly reduces the delay caused by beam scans.
\subsubsection{The blockage-prediction module}
The input of the blockage-prediction module can be written as
 \begin{eqnarray}\label{equfusing}
\bm x_{\{\textrm{bl},X_{\textrm{bm}}\}}=[f_{\{\textrm{en,a}\}}(X_{\textrm{bl},1}), f_{\{\textrm{en,sbl}\}} (\{X_{\textrm{bl},v}\}_{v=2}^{V_{\textrm{bl}}})].
\end{eqnarray}
With  $\bm x_{\{\textrm{bl},X_{\textrm{bl}}\}}$ as the input, the blockage-prediction module is composed of several FC, dropout, and Relu   layers. Define the  mathematical  function of the  blockage-prediction module as $f_{\{\textrm{dec,bl}\}}$.
Then, the output of the blockage-prediction module is
\begin{eqnarray}\label{equdecf}
y_{\{\textrm{bl},X_{\textrm{bl}}\}} =\sigma( f_{\{\textrm{dec,bl}\}}(\bm x_{\{\textrm{bl},X_{\textrm{bl}}\}})),
\end{eqnarray}
where $\sigma(z)=1/(1+e^{-z})$ is the sigmoid function.
We adopt the binary cross entropy loss for blockage prediction, which  is given by
\begin{align}\label{equdblockghg}
f_{\{\textrm{Lbl},X_{\textrm{bl}}\}}&=
 y_{\{\textrm{bl,lab}\}}   \log y_{\{\textrm{bl},X_{\textrm{bl}}\}}   \nonumber \\
&{\kern 30pt} + \left(1-y_{\{\textrm{bl,lab}\}}\right)   \log \left(1- y_{\{\textrm{bl},X_{\textrm{bl}}\}} \right),
\end{align}
where $y_{\{\textrm{bl,lab}\}}$ is the blockage label.
The auxiliary, the semantic-blockage, and the  blockage-prediction modules  are jointly trained by   ADAM  algorithm  to minimize $f_{\{\textrm{Lbl},X_{\textrm{bl}}\}}$ until the convergence.
In the testing stage, the blockage accuracy can be obtained by
\begin{eqnarray}\label{equaccblcok}
A_{\{\textrm{bl},X_{\textrm{bl}}\}}=\sum\nolimits_{n}^{N_{\textrm{tot}}}\mathds{1}(y_{\{\textrm{bl},X_{\textrm{bl}}\}} ^{(n)}=y_{\{\textrm{bl,lab}\}}^{(n)})/N_{\textrm{tot}}.
\end{eqnarray}
\subsubsection{FS for beam   prediction}

For simplicity, we adopt the Top-1 beam prediction accuracy as the optimization objective of FS.
Based on Eq.~\eqref{equacckdcn},  the Top-1 beam prediction accuracy with the feature set $X$ can be calculated by
\begin{eqnarray}\label{equacc}
A_{\{1,X_{\textrm{bm}}\}}=\sum\nolimits_{n}^{N_{\textrm{tot}}}\mathds{1}( y_{\{1,X_{\textrm{bm}}\}}^{(n)}= y_{\{\textrm{bm,lab}\}}^{(n)})/N_{\textrm{tot}},
\end{eqnarray}
where $y_{\{1,X_{\textrm{bm}}\}}^{(n)}$ is the  predicted Top-1 beam of the $n$-th sample, i.e.,
\begin{eqnarray}\label{equrep}
y_{\{1,X_{\textrm{bm}}\}}^{(n)}=\underset{m=0,\cdots,M_{\textrm{bm}}-1}{\arg \max } \ \bm y_{\{\textrm{bm},X_{\textrm{bm}}\}} ^{(n)}[m] .
\end{eqnarray}
Originated from the  SFFS algorithm, the proposed FS algorithm  takes the  universal feature set $X_{\textrm{uni}}$ as the input and then outputs the selected subset  $X_{\textrm{bm}}$, where the mapping mechanism from  the feature set  $X_{\textrm{bm}}$ to  the   Top-1 beam prediction accuracy $A_{\{1,X_{\textrm{bm}}\}}$ serves as the optimization objective.
Denote the selected subset at the $i$-th iteration  as $X_{\textrm{bm}}^{(i)}$.
The history feature set $X_{\{\textrm{bm,hist}\}}$ contains the history iterations  of the selected subset.
We initialize both  $X_{\textrm{bm}}^{(0)}$ and $X_{\{\textrm{bm,hist}\}}$  as  empty sets.
The  FS algorithm   can be divided into two alternate steps, i.e., {Step 1} and {Step 2}, until a termination criterion is satisfied.
 {Step 1}  aims to find the most significant feature $x^{+}$ among the unselected features, i.e., the feature leading to the best accuracy increase for  $X_{\textrm{bm}}^{(i)}$:
\begin{equation}\label{equxxplus}
 x^{+} \leftarrow \underset{x \in X_{\textrm{uni}}-X_{\textrm{bm}}^{(i)} }{\arg\max} A_{\{1,X_{\textrm{bm}}^{(i)}+x\}}.
\end{equation}
After successively updating $X_{\textrm{bm}}^{(i+1)}$, $X_{\{\textrm{bm,hist}\}}$, and $i$  by
\begin{align}\label{equxxplusv}
X_{\textrm{bm}}^{(i+1)} &\leftarrow X_{\textrm{bm}}^{(i)} + x^{+},\\
X_{\{\textrm{bm,hist}\}}& \leftarrow \{X_{\{\textrm{bm,hist}\}},X^{(i)}\},
\end{align}
and $ i \leftarrow i+1$, respectively,
we go over to {Step 2} with the goal to remove the least significant feature in $X_{\textrm{bm}}^{(i)}$.
In  {Step 2}, if there exists one feature in $X_{\textrm{bm}}^{(i)}$ that has a negative contribution to the accuracy, i.e.,  satisfying
\begin{equation}\label{equxx2if}
\underset{x \in   X_{\textrm{bm}}^{(i)} }{\max} A_{\{1,X_{\textrm{bm}}^{(i)}-x\}}  >A_{\{1,X_{\textrm{bm}}^{(i)}\}} ,
\end{equation}
then we will remove the feature $ x^{-}$ that results in the maximum accuracy drop, i.e., obtaining $ x^{-}$ by
\begin{equation}\label{equmina}
x^{-} \leftarrow \underset{x \in  X_{\textrm{bm}}^{(i)} }{\arg\max} A_{\{1,X_{\textrm{bm}}^{(i)}-x\}}.
\end{equation}
\begin{algorithm}[!t]
    \LinesNumbered
    \algsetup{linenosize=\tiny}  \small
    \caption{FS algorithm for beam   prediction }
    \label{notr}
    \KwIn{   Universal feature set $X_{\textrm{uni}}\buildrel \Delta \over =\{X_{\textrm{uni},v}| i=1,\cdots,V_{\textrm{uni}} \}$}
     \KwOut {Selected feature set $X_{\textrm{bm}}\buildrel \Delta \over =\{X_{\textrm{bm},v}| i=1,\cdots,V_{\textrm{bm}},X_{\textrm{bm},v} \in X_{\textrm{uni}}  \}$}
      Initialize feature set $X_{\textrm{bm}}^{(0)} \leftarrow \emptyset $ \\
      history feature set $X_{\{\textrm{bm,hist}\}}\leftarrow  \emptyset    $; $i=0$\\
    \textbf{Step 1: (Inclusion)} \\
    \quad\  \textbf{if} {$X_{\textrm{bm}}^{(i)}  \notin X_{\{\textrm{bm,hist}\}}$} \textbf{then} \label{linedff}\\
    \quad\  \quad\  $x^{+} \leftarrow \underset{x \in X_{\textrm{uni}}-X_{\textrm{bm}}^{(i)} }{\arg\max} A_{\{1,X_{\textrm{bm}}^{(i)}+x\}}$; $X_{\textrm{bm}}^{(i+1)} \leftarrow X_{\textrm{bm}}^{(i)} + x^{+}  $\\
    \quad\  \quad\  $X_{\{\textrm{bm,hist}\}}\leftarrow \{X_{\{\textrm{bm,hist}\}},X_{\textrm{bm}}^{(i)}\} $ \\
    \quad\  \quad\   $ i \leftarrow i+1$\\
    \quad\ \quad\   \textbf{go to Step 2}\\
    \quad\ \textbf{else}\\
    \quad\ \quad\   \textbf{Output} $ X_{\textrm{bm}} \leftarrow X_{\textrm{bm}}^{(i)}  $\\
    \quad\ \textbf{end}\\
    \textbf{Step 2: (Conditional Exclusion)} \\
    \quad\  \textbf{if} {$\underset{x \in  X_{\textrm{bm}}^{(i)} }{\max} A_{\{1,X_{\textrm{bm}}^{(i)}-x\}}  >A_{\{1,X_{\textrm{bm}}^{(i)}\}}$} \textbf{then} \\
   {
      \quad\ \quad\  $x^{-} \leftarrow \underset{x \in  X_{\textrm{bm}}^{(i)} }{\arg\max}A_{\{1,X_{\textrm{bm}}^{(i)}-x\}} $;  $X_{\textrm{bm}}^{(i+1)} \leftarrow X_{\textrm{bm}}^{(i)} - x^{-}  $ \\
      \quad\ \quad\   $ i \leftarrow i+1$\\
      \quad\ \quad\   \textbf{go to Step 2}\\
      \quad\ \textbf{else}\\
      \quad\ \quad\   \textbf{go to Step 1}\\
      \quad\ \textbf{end}\\
    }
\end{algorithm}
After   successively updating $ X_{\textrm{bm}}^{(i+1)}$   and $i$  by
\begin{equation}\label{equxxpddlusv}
X_{\textrm{bm}}^{(i+1)} \leftarrow X_{\textrm{bm}}^{(i)} - x^{-}
\end{equation}
and $ i \leftarrow i+1$,  we go back to   check the condition  Eq.~\eqref{equxx2if}.
If Eq.~\eqref{equxx2if} is not  satisfied, then it indicates any feature in $X_{\textrm{bm}}^{(i)}$ has a  positive contribution to the accuracy. In this case, we will come to  {Step 1} to select the most significant feature $x^{+}$ in the feature set $X_{\textrm{uni}}-X_{\textrm{bm}}^{(i)} $.
In  {Step 1}, when  $X_{\textrm{bm}}^{(i)}$ has appeared in  history feature set $X_{\{\textrm{bm,hist}\}}$, the iteration will enter an infinite same loop, i.e., the feature added in  {Step 1}  will be removed in  {Step 2}, which means we cannot find a feature set that outperforms $X_{\textrm{bm}}^{(i)}$ in terms of the beam prediction accuracy. Therefore, the algorithm ends and outputs   $ X_{\textrm{bm}} \leftarrow X_{\textrm{bm}}^{(i)}  $.
 The detailed steps of the FS algorithm  for beam prediction are given in   \textbf{Algorithm~\ref{notr}}.
In fact, other termination criterion could also be adopted. For instance, to limit the number of select features, i.e., $|X_{\textrm{bm}}|$ being no more   than $V_{\textrm{max}}$, we could replace   Line~\ref{linedff} of   \textbf{Algorithm~\ref{notr}} with ``$
\textbf{if} \quad X_{\textrm{bm}}^{(i)} \notin X_{\{\textrm{bm,hist}\}}\ \textbf{and} \ |X_{\textrm{bm}}^{(i)}|< V_{\textrm{max}}  \quad \textbf{then}$''.
In this way,  \textbf{Algorithm~\ref{notr}} can output the optimal feature subset $ X_{\textrm{bm}} $ with specified  size.

\subsubsection{FS for blockage prediction}
The FS algorithm for blockage prediction aims to  select  the  optimal  feature set $X_{\textrm{bl}}$. The detailed steps of the FS algorithm for blockage prediction
is the same with that for beam prediction except that  its optimization objective is replaced by  $A_{\{\textrm{bl},X_{\textrm{bl}}\}}$, as shown in  \textbf{Algorithm~\ref{notrnlck}}.
\begin{algorithm}[!t]
    \LinesNumbered
    \algsetup{linenosize=\tiny}  \small
    \caption{FS algorithm for blockage   prediction }
    \label{notrnlck}
    \KwIn{   Universal feature set $X_{\textrm{uni}}\buildrel \Delta \over =\{X_{\textrm{uni},v}| i=1,\cdots,V_{\textrm{uni}} \}$}
     \KwOut {Selected feature set $X_{\textrm{bl}}\buildrel \Delta \over =\{X_{\textrm{bl},v}| i=1,\cdots,V_{\textrm{bl}},X_{\textrm{bl},v} \in X_{\textrm{uni}}  \}$}
      Initialize feature set $X_{\textrm{bl}}^{(0)} \leftarrow \emptyset $\\
       history feature set $X_{\{\textrm{bl,hist}\}}\leftarrow  \emptyset    $;  $i=0$\\
    \textbf{Step 1: (Inclusion)} \\
    \quad\  \textbf{if} {$X_{\textrm{bl}}^{(i)}  \notin X_{\textrm{hist}}$} \textbf{then} \label{linedff}\\
    \quad\  \quad\  $x^{+} \leftarrow \underset{x \in X_{\textrm{uni}}-X_{\textrm{bl}}^{(i)} }{\arg\max} A_{\{\textrm{bl},X_{\textrm{bl}}^{(i)}+x\}}$;  $X_{\textrm{bl}}^{(i+1)} \leftarrow X_{\textrm{bl}}^{(i)} + x^{+}  $ \\  \quad\  \quad\   $X_{\{\textrm{bl,hist}\}}\leftarrow \{X_{\{\textrm{bl,hist}\}},X_{\textrm{bl}}^{(i)}\} $  \\
       \quad\  \quad\   $ i \leftarrow i+1$\\
    \quad\ \quad\   \textbf{go to Step 2}\\
    \quad\ \textbf{else}\\
    \quad\ \quad\   \textbf{Output} $ X_{\textrm{bl}} \leftarrow X_{\textrm{bl}}^{(i)}  $\\
    \quad\ \textbf{end}\\
    \textbf{Step 2: (Conditional Exclusion)} \\
    \quad\  \textbf{if} {$\underset{x \in  X_{\textrm{bl}}^{(i)} }{\max} A_{\{\textrm{bl},X_{\textrm{bl}}^{(i)}-x\}}  >A_{\{\textrm{bl},X_{\textrm{bl}}^{(i)}\}}$} \textbf{then} \\
   {
      \quad\ \quad\  $x^{-} \leftarrow \underset{x \in  X_{\textrm{bl}}^{(i)} }{\arg\max}A_{\{\textrm{bl},X_{\textrm{bl}}^{(i)}-x\}} $;  $X_{\textrm{bl}}^{(i+1)} \leftarrow X_{\textrm{bl}}^{(i)} - x^{-}  $\\
      \quad\  \quad\   $ i \leftarrow i+1$\\
      \quad\ \quad\   \textbf{go to Step 2}\\
      \quad\ \textbf{else}\\
      \quad\ \quad\   \textbf{go to Step 1}\\
      \quad\ \textbf{end}\\
    }
\end{algorithm}

\subsubsection{The overall training steps of the environment semantics aided
network architecture}
The overall implementation steps can be given  as follows:
\begin{enumerate}[i)]
  \item Train the environment semantics extraction network at each camera end, i.e, using ADAM algorithm to minimize Eq.~\eqref{equdentsemp} until the convergence. The extraction accuracy of the environment semantics is calculated by Eq.~\eqref{equaccpsp}.
  \item Apply \textbf{Algorithm~\ref{notr}} to select the optimal feature set $X_{\textrm{bm}}$ from   the universal feature set $X_{\textrm{uni}}$. For each iterated feature set, the  auxiliary, the semantic-beam, and the beam-prediction modules would be jointly retrained by minimizing Eq.~\eqref{equdentrop} until the convergence. The Top-1 beam prediction accuracy corresponding to the iterated feature set is tested by Eq.~\eqref{equacc}.
  \item Save and fix the parameters of the  auxiliary, the semantic-beam, and the beam-prediction modules corresponding to the optimal feature set $X_{\textrm{bm}}$.
  \item Apply \textbf{Algorithm~\ref{notrnlck}} to select the optimal feature set $X_{\textrm{bl}}$ from   the universal feature set $X_{\textrm{uni}}$. For each iterated feature set, the  auxiliary, the semantic-blockage, and the blockage-prediction modules would be jointly retrained by minimizing Eq.~\eqref{equdblockghg} until the convergence. The blockage accuracy corresponding to the iterated feature set is tested by Eq.~\eqref{equaccblcok}.
  \item Save and fix the parameters of the  auxiliary, the semantic-blockage, and the blockage-prediction modules corresponding to the optimal feature set $X_{\textrm{bl}}$. Note that the   initialization,   the training, and the storage of the auxiliary module in Step iv)-Step v) are all independent with those in Step ii)-Step iii).
  \item The cameras only send the selected environment semantics, i.e., $X_{\textrm{bm}}\cup X_{\textrm{bl}}$,  to the task-oriented encoder. Then,  BS obtains the predicted Top-$G$ beam and blockage by the decision network.
\end{enumerate}

\textbf{Remark:}
Compared with the pervious vision based works \cite{weihua20192d,9523557,9512383} that directly utilize environment image data to assist channel related tasks, the  environment semantics aided network architecture enjoys two main benefits:
\begin{itemize}
  \item Environment semantics are  extracted  from environment images for subsequent data processing, which can not only  protect  user privacy especially in the case of using the third part cameras, but also reduce  the system overheads (e.g., storage space and computational cost) brought by channel-irrelevant information.
  \item Channel semantics is extracted from environment images, which reveals the  key scatterers associated with  channel propagation paths and further provides an interpretable insight into the joint communication and sensing systems.
\end{itemize}

\section{Simulation Results}\label{secsimu}
\subsection{Simulation  Setup}

\subsubsection{Environment modeling}
The autonomous driving simulator CARLA \cite{Dosovitskiy17} is utilized to simulate the  street traffic environment,
including the street landscape, vehicles,  and   cameras, etc.
Vehicles  in the deployment scenario are randomly chosen from the three types, i.e.,    the car (3.71$\times$1.79$\times$1.55m$^3$), the van (5.20$\times$2.61$\times$2.47m$^3$), and the bus (11.08$\times$3.25$\times$3.33m$^3$).
Then, we utilize the traffic simulation software SUMO \cite{SUMO2018}  to control the speed and
trajectory of all the vehicles.
A partial view of the street traffic environment is shown in the left of Fig.~\ref{fignetmodel}.

 \subsubsection{Scene image and semantic label generation} \label{secsema}
As shown in Fig.~\ref{fignetmodel}, two cameras  are  set at 5m on both sides of the street, both orienting towards the street. The two cameras  keep taking images  as  vehicles  cross the street.
Fortunately, the CARLA simulator can generate the corresponding semantic label for each images taken by cameras.
There are 20 classes of environmental semantic concepts, including ``building'', ``fence'', ``pedestrian'', ``pole'', ``roadline'',
 ``sidewalk'', ``vegetation'', ``vehicle'', ``wall'', ``trafficsign'', ``sky'', ``ground'', ``bridge'', ``railtrack'', ``trafficlight'', ``static'', ``dynamic'', ``water'', ``terrain'', and ``unlabeled''.
In particular, the concept ``static'' refers to the elements  that are immovable, e.g., fire hydrants, fixed benches, and bus stops, etc. The concept ``dynamic'' refers to the elements that are susceptible to move over time, e.g.,  wheelchairs, animals, and  buggies, etc.
The concept ``unlabeled'' refers to the elements that have not been categorized.
By collecting  the \{image, semantic label\}  pairs  and splitting them to the training and the testing datasets,
 the  semantic feature extraction network PSPNet \cite{8100143} could be trained to produce the corresponding environment semantics.

 \begin{figure}[!t]
\centering
\includegraphics[width=65mm]{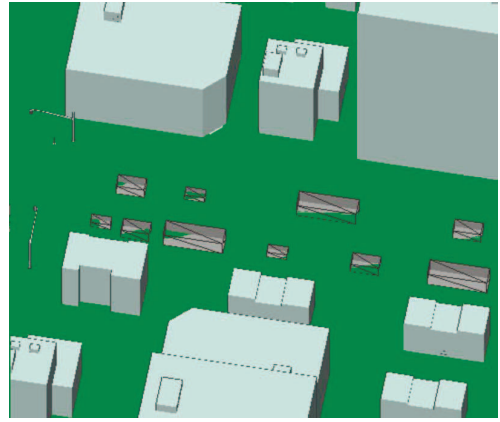}
\caption{A partial view  of the synchronization simulation in the Wireless Insite simulator.
To save the cost of art designs, we utilize simple cubes to model  the buildings and the vehicles in Wireless Insite and ignore the surface details of these elements. The losses in surface details  have limited influences on the channels, which will not affect the reliability of simulations.}
\label{figinste}
\end{figure}
\begin{table*}[!t]\small
\centering
\caption{The network parameters for the   beam prediction and blockage prediction}
\label{tabsfgder}
\begin{tabular}{|c|c|c|c|c|c|  }
\hline
\footnotesize Module & \footnotesize Layer   & \footnotesize Kernel  & \footnotesize Stride & \footnotesize Filter/Neuron    &\footnotesize Output Shape\\
 \hline
 \multirow{2}[2]{*}{ \makecell[c]{\footnotesize Auxiliary}}  &BatchNorm     & /  & /  & 3 &   3    \\
  &(FC-BatchNorm-ReLu)$\times$2   & /  & /  & 256,16 &   16 \\
\hline
 \multirow{4}[2]{*}{ \makecell[c]{\footnotesize Semantic-beam}}
  &(Conv-BatchNorm-ReLu)$\times$2 &  3, 3     &  4, 2   & 32, 16 &   16$\times$40$\times$80  \\
  & AvgPool &  3     &  2   & / &   16$\times$20$\times$40   \\
  &(Conv-BatchNorm-ReLu)$\times$2$<$Residual$>$   & 3, 3   &  4, 1  & 8, 8 &   8$\times$5$\times$10 \\
  &(Conv-BatchNorm-ReLu)$\times$2$<$Residual$>$  & 3, 3   &  1, 1  & 8, 8 &   8$\times$5$\times$10 \\
\hline
 \multirow{3}[2]{*}{ \makecell[c]{\footnotesize Beam-prediction}}  &(FC-BatchNorm-ReLu)     & /  & /  & 512  &   512   \\
 \cline{3-5}
  & Dropout    & \multicolumn{3}{c|}{Rate: 0.1} &   512 \\
   \cline{3-5}
    & FC    & /  & /  & 64  &       64 \\
\hline
 \multirow{3}[2]{*}{ \makecell[c]{\footnotesize Semantic-blockage}}
  &(Conv-BatchNorm-ReLu) &   3     &   2   & 16 &   16$\times$40$\times$80  \\
  & AvgPool &  3     &  2   & / &   16$\times$20$\times$40   \\
  &(Conv-BatchNorm-ReLu)$\times$2$<$Residual$>$   & 3, 3   &  4, 1  & 8, 8 &   8$\times$5$\times$10 \\
\hline
 \multirow{3}[2]{*}{ \makecell[c]{\footnotesize Blockage-prediction}}  &(FC-BatchNorm-ReLu)     & /  & /  & 64  &   64   \\
 \cline{3-5}
  & Dropout    & \multicolumn{3}{c|}{Rate: 0.1} &   64 \\
   \cline{3-5}
    & FC    & /  & /  & 1  &       1 \\
\hline
\end{tabular}
\end{table*}

\subsubsection{Channel, blockage label, and beam label generation}
To obtain the channels of users in  vehicles, we  synchronize the street traffic environment in CARLA with
the  3D ray-tracing simulator Wireless  InSite \cite{alkhateeb2019deepmimo} at each moment that cameras take images, which ensures that the images taken by the two cameras exactly correspond  to the  channels.
Since one image may contain several vehicles or users, we also record the location  of the target user, to form the uniquely identified sample pair of the two environment images, the location, and the channel.
Fig.~\ref{figinste} displays a  partial view  of the synchronization simulation in the Wireless Insite simulator, where
the  buildings and the sizes/locations/orientations of vehicles in Wireless Insite are exactly the same as that in CARLA at each image shooting  moment. The carrier frequency is 28 GHz, the number of OFDM subcarriers is 128, and  the BS is equipped with 64 antennas.
After setting the transceiver antennas at the corresponding locations, the ray-tracing simulator shoots thousands of rays in all directions from the transmitter and records the strongest 20 paths that reach the receiver,  obtaining  the corresponding channel parameters, i.e., $\{\alpha _{ l},\phi_{ l},\tau_{ l},\theta_{\mathrm{az}}^{ l}, \theta_{\mathrm{el}}^{ l}\}$, and further obtaining  the channels using   Eq.~\eqref{equchannel}.
The  blockage labels can be obtained according to whether there exists a LOS path among the   strongest 20 paths during the future time period between two  image shootings. Each time slot is 50ms. Unless otherwise specified, the blockage prediction is to predict the next future time slot, i..e., blockage state in next 50ms.
Following Eq.~\eqref{equaraterhh}, the corresponding beam labels can be obtained by  exhaustively  searching  on the codebook. Here we adopt the discrete fourier transformation codebook as an example.

\subsubsection{Network parameters}
The network structure of  the feature extraction network follows the work \cite{8100143}, except that the input and the output dimensions are adjusted according to the shapes of the images and the semantic labels generated by CARLA. Moreover, various network scales, including ResNet-18, ResNet-34, ResNet-50, and ResNet-100 \cite{7780459},  are respectively adopted and trained on  the datasets  in Section~\ref{secsema}.
The network parameters for the   auxiliary, the semantic-beam, the semantic-blockage, the beam-prediction, and the blockage-prediction modules are given in Tab.~\ref{tabsfgder}, where
``-'' represents sequential stacking,  ``$(\cdot)\times 2$'' represents that the layers insides $(\cdot)$ are sequentially stacked twice, the numbers split by ``,'' represents the parameters for  adjacent convolution or FC layers, ``$<$Residual$>$'' represents the residual connection between the  adjacent convolutional layers, and  ``Conv'' is the abbreviation of ``convolution''.
The initial learning rate of the ADAM optimizer is 0.001, and the batch size is 128.

  \begin{figure}[!t]
\centering
\includegraphics[width=65 mm]{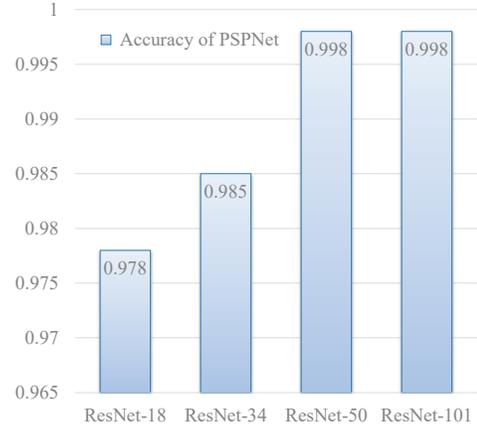}
\caption{The prediction accuracy of  PSPNet, i.e., the accuracy of the environment semantics extraction, trained on datasets  in Section~\ref{secsema}.}
\label{figpspggr}
\end{figure}
 \begin{figure*}
\centering     
\subfigure[Top-5 beam prediction accuracy versus the number of selected features.]{\label{figfeaff1}\includegraphics[width=75mm]{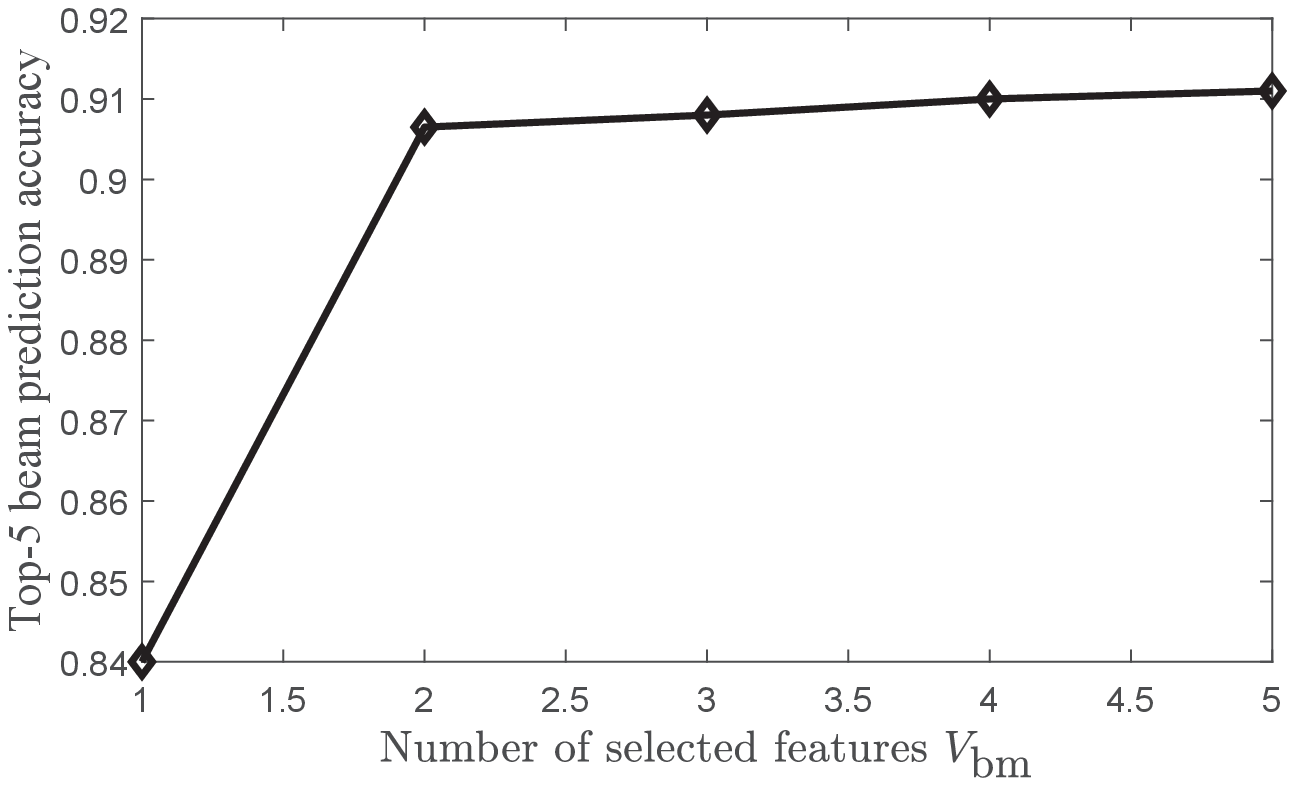}}
\subfigure[Top-5 beam prediction accuracy under  various selected features, $V_{\textrm{bm}} =1$.]{
\label{figfeaff2}\includegraphics[width=75mm]{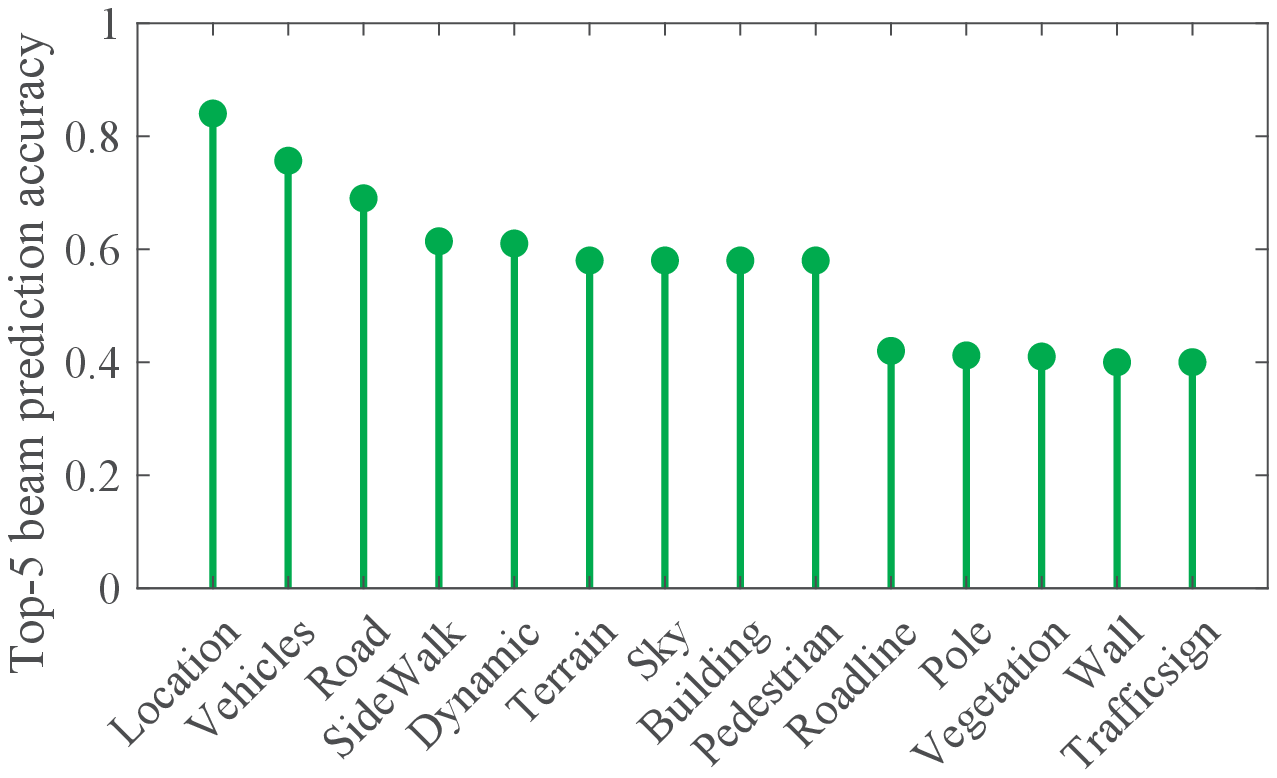}
}\\
\subfigure[Top-5 beam prediction accuracy under  various selected features, $V_{\textrm{bm}} =2$.]{
\label{figfeaff3}\includegraphics[width=75mm]{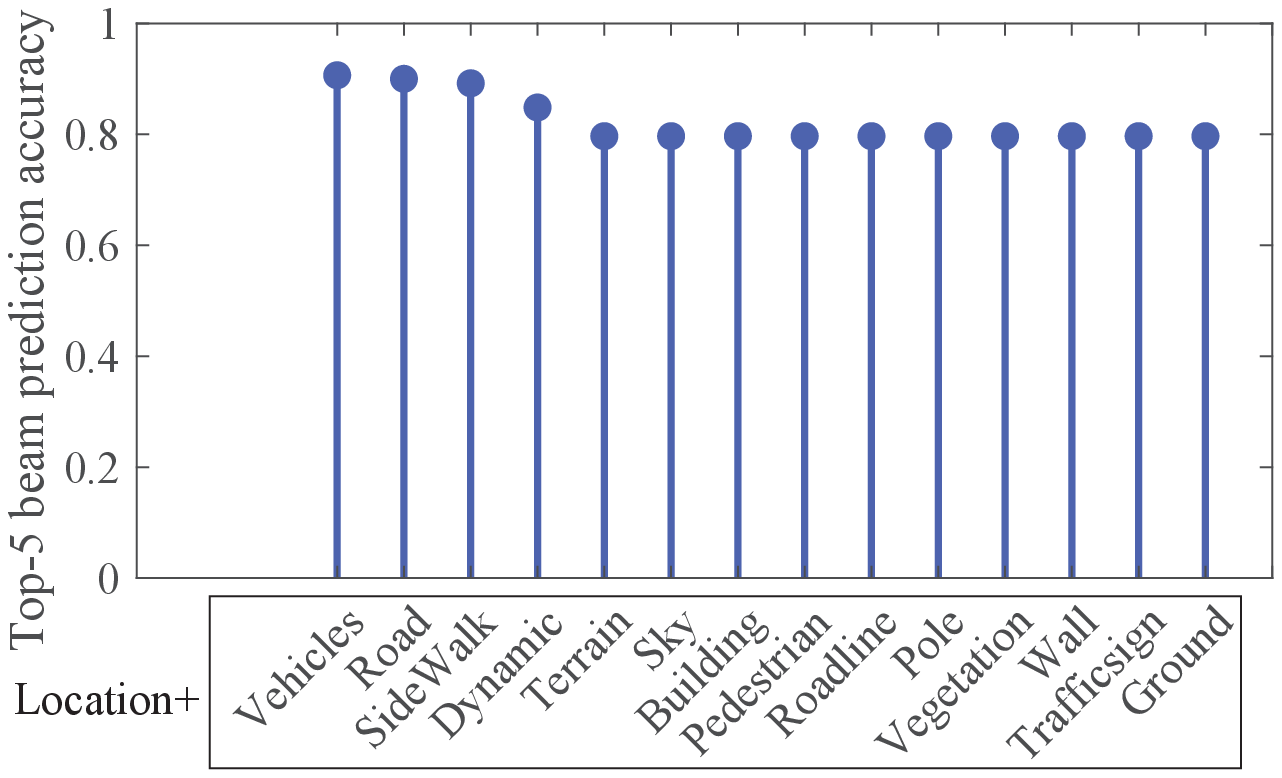}
}
\subfigure[Top-5 beam prediction accuracy under  various selected features, $V_{\textrm{bm}} =3$.]{\label{figfeaff4}\includegraphics[width=75mm]{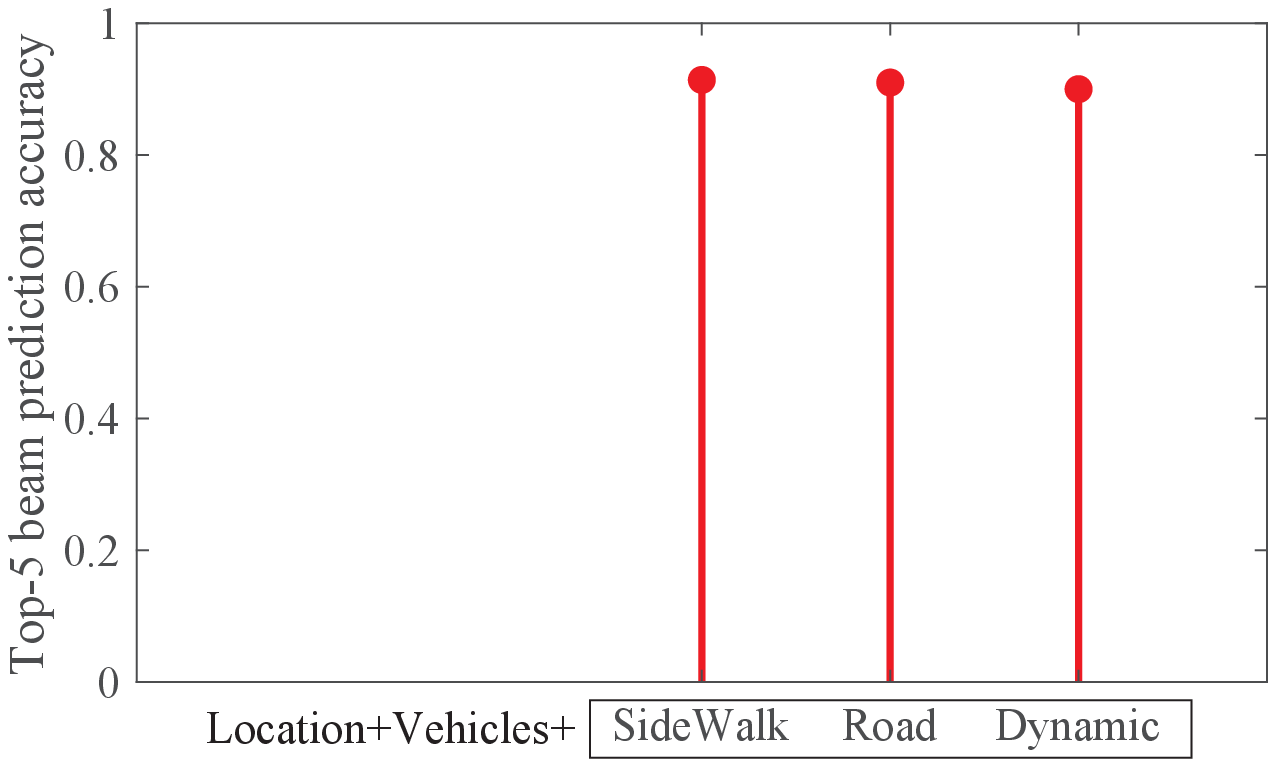}}
\caption{The Top-5 beam prediction accuracy $A_{\{5,X_{\textrm{bm}}\}}$ under various selected  feature sets.}
\label{figfeatureselect}
\end{figure*}
\subsection{Performance of the Environment Semantics  Extraction}
We adopt PSPNet to extract the environment semantics  from environment images, which implies that the  prediction accuracy of  PSPNet, i.e., $A_{\textrm{PSP}}$, is the accuracy of the environment semantics extraction.
Fig.~\ref{figpspggr} exhibits the prediction accuracies of  PSPNet   versus the network scales.
The accuracy of PSPNet improves as the network scale increases, and the accuracy  saturates when the network scale reaches that of ResNet-50.
The work \cite{8100143} presents the  accuracy of PSPNet on  the ADE20K dataset \cite{ZhougdtBolei}, which is much lower than that on
the datasets generated by CARLA.
This is because that the  number  of semantic concepts to be classified in the ADE20K dataset is  150, which is much larger than that in  our datasets, i.e., 20. Therefore, the classification task  for the ADE20K dataset is harder than  that for  our datasets.
 This indicates that the accuracy of the semantic feature extraction could  be further improved if we could remove unimportant  semantic features and further reduce the number  of categories to be classified.

\subsection{Performance of FS for Beam Prediction and Blockage Prediction}

Fig.~\ref{figfeatureselect} displays the Top-5 beam prediction accuracy $A_{\{5,X_{\textrm{bm}}\}}$ under various  feature sets selected by \textbf{Algorithm~\ref{notr}}. Fig.~\ref{figfeaff1} plots the Top-5 beam prediction accuracy versus the number of selected features $V$, where the Top-5 beam prediction accuracy is obtained through the execution of \textbf{Algorithm~\ref{notr}}.
More specifically,   the Top-5 beam prediction accuracy at $V_{\textrm{bm}} =v$   can be determined by
\begin{equation*}
\underset{X_{\textrm{bm}} \in X_{\{\textrm{bm,hist}\}} ,|X_{\textrm{bm}}|=v } {\arg\max} A_{\{5,X_{\textrm{bm}}\}}.
\end{equation*}
The Top-5 beam prediction accuracy improves as $V_{\textrm{bm}} $ increases, while the   accuracy gain is less than 0.01 when $V_{\textrm{bm}} $ is larger than 3, which implies that three features are efficient enough for beam prediction.
Fig.~\ref{figfeaff2}-Fig.~\ref{figfeaff4}  present the Top-5 beam accuracies under  various selected  feature sets with $V_{\textrm{bm}} =1,2,$ and 3, respectively, where the feature names are listed in descending order along the $x$-axis  according to the corresponding accuracies.
In Fig.~\ref{figfeaff2}, only one feature is utilized to predict the beam.
The location yields the highest accuracy among all the features. This is because that the location is the  user's
identification information, which is essential to distinguish the target user from  other possible users\footnote{The accuracy performance associated with the location also explains that  location based works are popular \cite{weihua20192d,9523557,9512383}. In fact, other user's identification information, such
as   historical beam indices  or other predetermined information,   could also be used to
distinguish the target user from  other possible users. Therefore, the location based prediction is only a special case of the environment  semantics aided communication system.  }.
The reason that other features like the vehicle still yield higher accuracies than random predictions is due to the limited concurrent users within the same image.
In Fig.~\ref{figfeaff3}, two features are used to predict the beam, and one of them is the location.
It can be seen than the three environment semantic concepts, i.e.,  ``vehicle'', ``road'', and ``sidewalk'',  lead  to the highest beam  prediction accuracy.
In Fig.~\ref{figfeaff4}, three features are utilized to predict the beam, and two of them are the location and the vehicle. The feature set \{location, vehicle, sidewalk\} can achieve the highest accuracy, which conforms to the intuition that the vehicles provide the information of  users' directions and the possible obstacles, while sidewalks  offer  the information of street layouts.

 \begin{figure*}
\centering     
\subfigure[Blockage prediction accuracy  versus the number of selected features.]{\label{figfeablock1}\includegraphics[width=75mm]{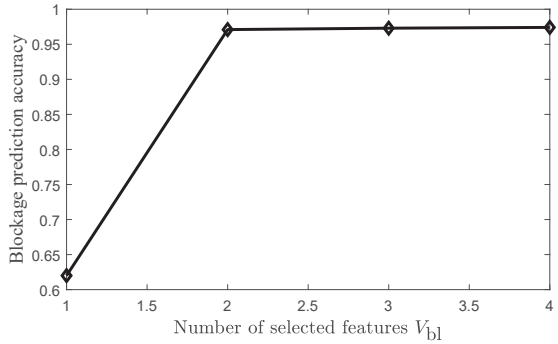}}
\subfigure[Blockage prediction accuracy  under  various selected features, $V_{\textrm{bl}}=1$.]{
\label{figfeablock2}\includegraphics[width=75mm]{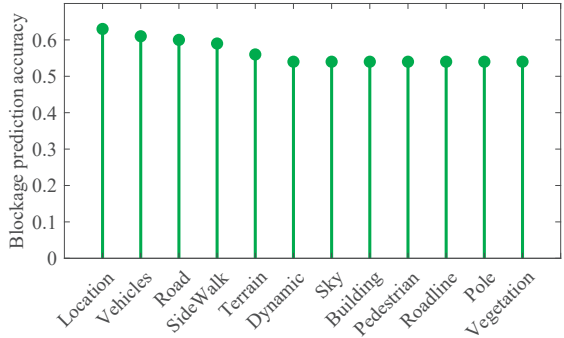}
}\\
\subfigure[Blockage prediction accuracy  under  various selected features, $V_{\textrm{bl}}=2$.]{
\label{figfeablock3}\includegraphics[width=75mm]{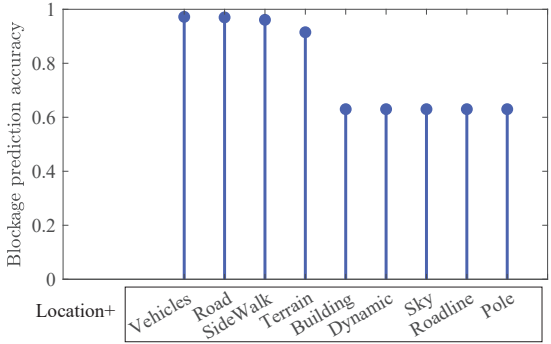}
}
\subfigure[Blockage prediction accuracy under  various selected features, $V_{\textrm{bl}}=3$.]{\label{figfeablock4}\includegraphics[width=75mm]{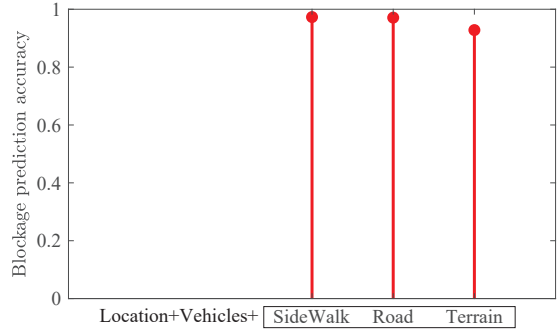}}
\caption{The blockage prediction accuracy $A_{\{\textrm{bl},X_{\textrm{bl}}\}}$  under various selected  feature sets.}
\label{figfeablock}
\end{figure*}

Fig.~\ref{figfeablock} displays the  blockage prediction accuracy $A_{\{\textrm{bl},X_{\textrm{bl}}\}}$  under various  feature sets selected by  \textbf{Algorithm~\ref{notrnlck}}.
Fig.~\ref{figfeablock1} plots the  blockage prediction accuracy versus the number of selected features $V_{\textrm{bl}}$, where the blockage prediction accuracy  at $V_{\textrm{bl}} =v$   can be obtained by
\[\underset{X_{\textrm{bl}} \in X_{\{\textrm{bl,hist}\}} ,|X_{\textrm{bl}}|=v } {\arg\max} A_{\{{\textrm{bl}},X_{\textrm{bl}}\}}.\]
It can be observed that two  features are efficient enough for blockage prediction.
Fig.~\ref{figfeablock2}-Fig.~\ref{figfeablock4}  present the  blockage prediction  accuracies under  various selected  feature sets with $V_{\textrm{bl}} =1,2,$ and 3, respectively, where the feature names are listed in descending order along the $x$-axis  according to the corresponding accuracies.
Similar with Fig.~\ref{figfeatureselect}, the feature set \{location, vehicle, sidewalk\} can achieve the highest accuracy.
This validates that the feature set \{location, vehicle, sidewalk\}  are
the most significant  features for channel related tasks, and other unselected semantic features
have lower relevancy with channels.

\begin{table}[!t]
  \centering
  \caption{Environment semantics analyses}
    \begin{tabular}{p{14em}p{8em}}
    \rowcolor[rgb]{ .267,  .447,  .769} \textcolor[rgb]{ 1,  1,  1}{\footnotesize \textbf{Environmental semantic concept}} & \textcolor[rgb]{ 1,  1,  1}{\footnotesize \textbf{Accuracy contribution}}   \\
        \hline
    \footnotesize Vehicles, Road, Sidewalk & \footnotesize Critical \\
        \hline
    \footnotesize Dynamic, Terrain  & \footnotesize Moderate \\
        \hline
    \footnotesize{Sky, Building, Pedestrian, Roadline, Pole, Vegetation, Wall, Ground, Bridge, Railtrack, Trafficlight, Water, Fence, Static} & \footnotesize \multirow{2}[1]{*}{Negligible} \\
        \hline
    \end{tabular}%
  \label{tabanattt}%
\end{table}%

\begin{table}[!t]
  \centering
  \caption{Beam prediction and blockage prediction accuracies with various inputs}
    \begin{tabular}{ccc}
      \hline
     \footnotesize {\textbf{Input}} & \footnotesize {\textbf{Top-5 beam}} & \footnotesize  {\textbf{Blockage}}   \\
        \hline
   \footnotesize Original  images & \footnotesize 0.910$\pm$0.091  & \footnotesize 0.971$\pm$0.088  \\
        \hline
   \footnotesize All environment semantics  &\footnotesize 0.914$\pm$0.042& \footnotesize 0.974$\pm$0.036 \\
        \hline
    \footnotesize \{Vehicles, Road, Sidewalk\} &\footnotesize 0.914$\pm$0.020 & \footnotesize 0.974$\pm$0.018  \\
 \hline
    \end{tabular}%
  \label{tabbeamblock}%
\end{table}%
\begin{figure*}
\centering     
\subfigure[Beam accuracy versus the number of training epoches.]{\label{figtopk1}\includegraphics[width=80mm]{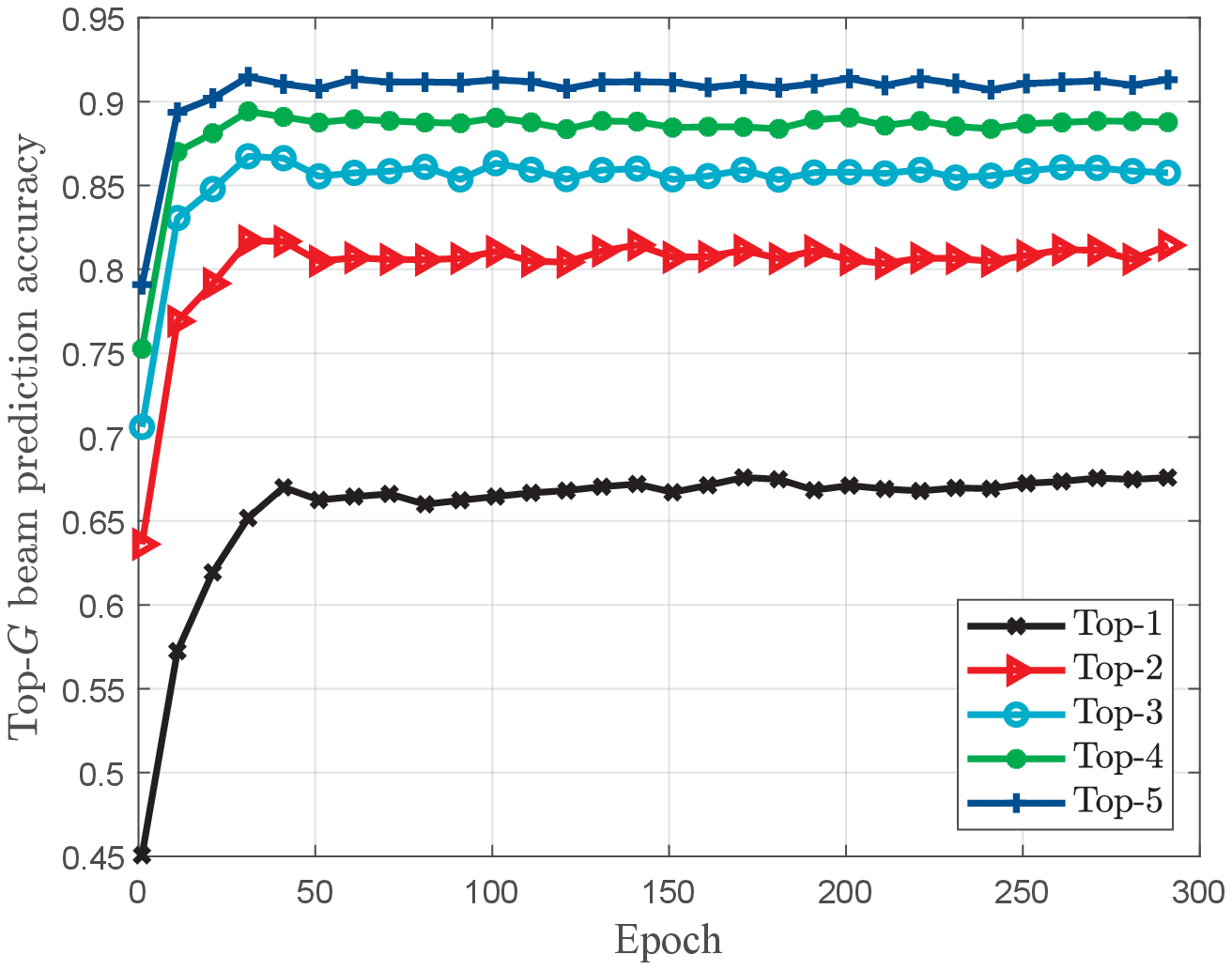}}
\subfigure[TRR versus the number of training epoches.]{
\label{figtopk2}\includegraphics[width=80mm]{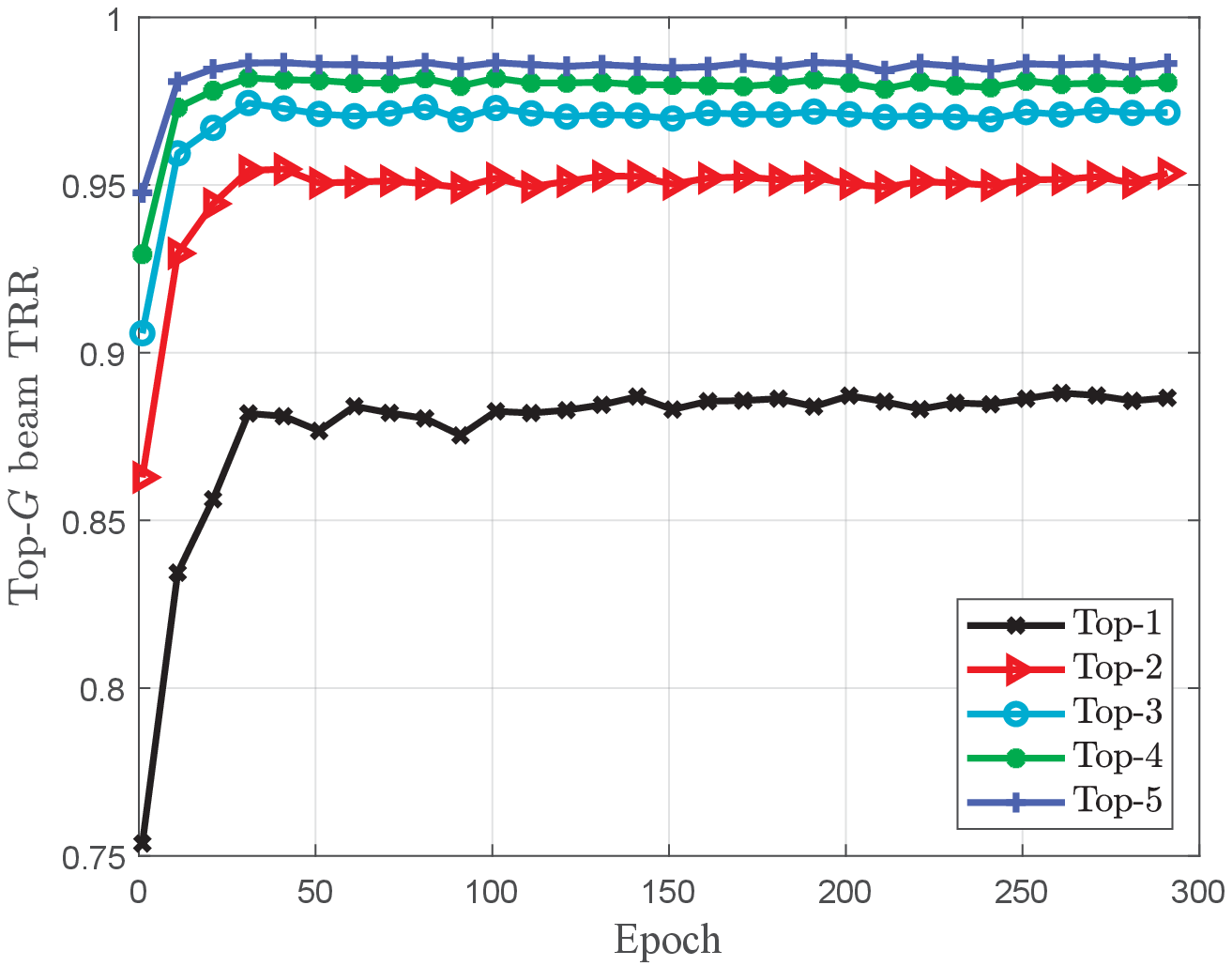}
}
\caption{The beam prediction accuracy (a) and  the TRR (b) of the Top-$G$ beam versus the number of training epoches.}
\label{figtopk}
\end{figure*}
\begin{table*}[!t]
  \centering
  \caption{Blockage prediction accuracy versus future time slots}
    \begin{tabular}{|c|c|c|c|c|c|c|c|c|}
    \hline
    \footnotesize Time slots &\footnotesize  1     &\footnotesize  6     &\footnotesize  11    & \footnotesize 16    & \footnotesize 21    & \footnotesize 26    & \footnotesize 31    & \footnotesize 36 \\
    \hline
    \footnotesize $A_{\{\textrm{bl},X_{\textrm{bl}}\}}$  &\footnotesize  0.9741 &\footnotesize  0.966 &\footnotesize   0.951 &\footnotesize  0.94  & \footnotesize 0.929 & \footnotesize 0.922 &\footnotesize  0.901 & \footnotesize 0.882 \\
    \hline
    \end{tabular}
  \label{tabggggfhjkd}
\end{table*}
Based on the results of Fig.~\ref{figfeatureselect} and Fig.~\ref{figfeablock}, we divide the environmental semantic concepts  into three levels of accuracy contributions, i.e., critical, moderate, and negligible, as shown in Tab.~\ref{tabanattt}.
Obviously, the environmental semantic concepts  ``vehicles'', ``road'', and ``sidewalk'' are the most critical features due to their contributions to the beam prediction and blockage prediction.
Therefore, the three environmental semantic concepts ``vehicles'', ``road'', and ``sidewalk'' can be interpreted as the most significant channel  semantics, and other unselected semantic features have lower relevancy with the channels.
For beam prediction, as shown in Fig.~\ref{figfeaff3} and Fig.~\ref{figfeaff4}, ``road'' is more effective than  ``sidewalk'' for $V_{\textrm{bm}}=2$ but is less  effective than  ``sidewalk'' for $V_{\textrm{bm}}=3$.
To understand the phenomenon, one can glean some indications in the right part of Fig.~\ref{fignetseg}, where the shapes of ``road'' and ``vehicles'' are complementary, which implies that the better accuracy of  ``road'' for  $V_{\textrm{bm}}=2$ owns to its  high dependency  on ``vehicles'', while the gain of  ``road'' would be weakened  for $V_{\textrm{bm}}=3$ where ``vehicles'' has been selected.
When ``vehicle'' has been selected, ``sidewalk'' could offer extra  beam prediction accuracy gains.
 Moreover, ``dynamic'' has moderate contributions to the beam prediction accuracy. This is because that  ``dynamic'' objects in the road may change  the shape of  ``vehicles'', and  the acquisition of ``dynamic'' can modify the shape distortion, yielding noticeable contributions to the beam prediction accuracy.
For blockage prediction, as shown in Fig.~\ref{figfeablock3} and Fig.~\ref{figfeablock4}, the accuracy gain brought by the third feature sidewalk is marginal. This indicates that the environmental semantic concept ``vehicles'' is sufficient for   blockage prediction, and blockage prediction is less sensitive to other environmental semantic concepts than beam prediction. This is because that the beam vector is highly related to various   scatterers in environments while the blockage state is only related to the  obstacle in the LOS path.

Tab.~\ref{tabbeamblock} compares the  beam prediction and blockage prediction accuracies with various  inputs, i.e., original images, all environment semantics, and the selected channel semantics set \{location, vehicle, sidewalk\}. The overall network structures for  original images and  all environment semantics are the same with those for the selected channel semantics, and the only difference is that the network input has been replaced by the original images and  the segmentation maps, respectively.
It can be observed that the channel semantics set is much sparse than the  environment semantics set, but is informative  enough to obtain almost the same prediction accuracy, which also demonstrates that the proposed FS  algorithm can significantly compress the original image flows while preserve the most informative features.
Furthermore, the standard deviations of the selected channel semantics set are smaller than those of original images and all environment semantics. This is because irrelevant semantics or fine image details expand the input feature space, making available samples sparser and more dissimilar, which results in a network  more easier to overfit and generate unstable predictions.
Therefore, the proposed environment semantics aided network architecture cannot only reduce system overheads  brought by channel-irrelevant information, but also improve  the prediction  robustness.

\subsection{Performance of Top-$G$ Beam Prediction}

Fig.~\ref{figtopk} depicts the beam prediction accuracy and  the TRR  of the Top-$G$ beam versus the number of training epoches, where the feature set \{location, vehicle, sidewalk\} is adopted.
It can be seen that the   task-oriented encoders and decision network converge  when the number of epoches is larger than 50.
As shown in Fig.~\ref{figtopk1}, the Top-1 beam prediction accuracy reaches 66\%,  and the Top-5 beam prediction accuracy reaches 91\%.
As shown in Fig.~\ref{figtopk2}, the TRR of Top-1 beam   reaches 88\%,  and the TRR of Top-5 beam  reaches 98\%.
These results demonstrate  the effectiveness of the  proposed  environment semantics aided network architecture.

\subsection{Performance of Blockage Prediction Versus Future Time Slots}
Tab.~\ref{tabggggfhjkd} displays the  blockage prediction accuracies versus future time slots, where each time slot is 50ms. For each recorded time slot point, i.e., 6, 12, $\cdots$, 36, the network would be retrained with the corresponding future blockage labels. It can be seen that the blockage prediction accuracy is higher than 90\% for future 30 time slots, i.e., 1.5s, which demonstrates the effectiveness of the proposed environment semantics aided network architecture.

\section{Conclusion} \label{secconcul}
In this paper, we propose a  framework for environment semantics aided wireless communications to improve the transmission efficiency and protect the user privacy.
 We also developed an  environment semantics aided network architecture for mmWave beam prediction and blockage prediction as  a case study. Simulation results show that only three features of the environment semantics can achieve almost the same accuracy as the whole environment semantics, which  can significantly reduce system overheads and improve the transmission efficiency.
 The superiority of  the environment semantics aided wireless communication framework in
 realizing extremely low-latency beam prediction and supporting ultra-reliable communication links
  demonstrates its great application potential  in URLLCs.

\bibliographystyle{IEEEtran}
\bibliography{References}


\end{document}